\documentclass[longauth]{aa}  

\usepackage{graphicx}
\usepackage{txfonts}
\usepackage{natbib}
\bibpunct{(}{)}{;}{a}{}{,} 
\newcommand{\ixpe}{{IXPE}\xspace}
\newcommand{\Msun}{{\it M}_{\odot}}

\usepackage{hyperref}
\newcommand{\orcidlink}[1]{\protect\href{https://orcid.org/#1}{\protect\includegraphics[width=8pt]{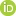}}}

\begin{document} 

\title{\ixpe observation confirms a high spin in the accreting black hole 4U~1957+115}

\author{
    L.~Marra\inst{\ref{I1}}\thanks{E-mail: lorenzo.marra@uniroma3.it}\orcidlink{0009-0001-4644-194X}\and 
    M.~Brigitte\inst{\ref{I2},\ref{I3},\ref{I4}} \orcidlink{0009-0004-1197-5935}\and
    N.~Rodriguez Cavero\inst{\ref{I5}} \orcidlink{0000-0001-5256-0278}\and 
    S.~Chun\inst{\ref{I5}} \orcidlink{0009-0002-2488-5272}\and
    J.~F.~Steiner\inst{\ref{I6}} \orcidlink{0000-0002-5872-6061}\and
    M.~Dov\v{c}iak\inst{\ref{I2}} \orcidlink{0000-0003-0079-1239}\and
    M.~Nowak\inst{\ref{I7}} \orcidlink{0000-0001-6923-1315}\and
    S.~Bianchi\inst{\ref{I1}} \orcidlink{0000-0002-4622-4240}\and
    F.~Capitanio\inst{\ref{I8}} \orcidlink{0000-0002-6384-3027}\and
    A.~Ingram\inst{\ref{I9}} \orcidlink{0000-0002-5311-9078}\and
    G.~Matt\inst{\ref{I1}} \orcidlink{0000-0002-2152-0916}\and
    F.~Muleri\inst{\ref{I8}} \orcidlink{0000-0003-3331-3794}\and
    J.~Podgorný\inst{\ref{I2},\ref{I3},\ref{I10}} \orcidlink{0000-0001-5418-291X}\and
    J.~Poutanen\inst{\ref{I11}}  \orcidlink{0000-0002-0983-0049}\and
    J.~Svoboda\inst{\ref{I2}}  \orcidlink{0000-0003-2931-0742}\and
    R.~Taverna\inst{\ref{I12}}  \orcidlink{0000-0002-1768-618X}\and
    F.~Ursini\inst{\ref{I1}}  \orcidlink{0000-0001-9442-7897}\and
    A.~Veledina\inst{\ref{I11},\ref{I13}}  \orcidlink{0000-0002-5767-7253}\and
    A.~De Rosa\inst{\ref{I8}}  \orcidlink{0000-0001-5668-6863}\and
    J.~A.~Garc\'{i}a\inst{\ref{I14}}  \orcidlink{0000-0003-3828-2448}\and
    A.~A.~Lutovinov\inst{\ref{IA}} \orcidlink{0000-0002-6255-9972}
    I.~A.~Mereminskiy\inst{\ref{IA}} \orcidlink{0000-0001-8931-1320}
    R.~Farinelli\inst{\ref{IB}} \orcidlink{0000-0003-2212-367X}
    S.~Gunji\inst{\ref{I15}}  \orcidlink{0000-0002-5881-2445}\and
    P.~Kaaret\inst{\ref{I16}}  \orcidlink{0000-0002-3638-0637}\and
    T.~Kallman\inst{\ref{I17}}  \orcidlink{0000-0002-5779-6906}\and
    H.~Krawczynski\inst{\ref{I5}}  \orcidlink{0000-0002-1084-6507}\and
    Y.~Kan\inst{\ref{I15}} \and
    K.~Hu\inst{\ref{I5}}  \orcidlink{0000-0002-9705-7948}\and
    A.~Marinucci\inst{\ref{I18}}  \orcidlink{0000-0002-2055-4946}\and
    G.~Mastroserio\inst{\ref{I19}}  \orcidlink{0000-0003-4216-7936}\and
    R.~Miku\u{s}incov\'a\inst{\ref{I1},\ref{I8}}  \orcidlink{0000-0001-7374-843X}\and
    M.~Parra\inst{\ref{I1},\ref{I20}}  \orcidlink{0009-0003-8610-853X}\and
    P.-O.~Petrucci\inst{\ref{I20}}  \orcidlink{0000-0001-6061-3480}\and
    A.~Ratheesh\inst{\ref{I8}}  \orcidlink{0000-0003-0411-4243}\and
    P.~Soffitta\inst{\ref{I8}}  \orcidlink{0000-0002-7781-4104}\and
    F.~Tombesi\inst{\ref{I23},\ref{I22},\ref{I21}} \orcidlink{0000-0002-6562-8654}\and
    S.~Zane\inst{\ref{I24}} \orcidlink{0000-0001-5326-880X}\and
    I.~Agudo\inst{\ref{IS1}} \orcidlink{0000-0002-3777-6182}\and
    L.~A.~Antonelli\inst{\ref{IS2},\ref{IS3}} \orcidlink{0000-0002-5037-9034}\and
    M.~Bachetti\inst{\ref{I19}} \orcidlink{0000-0002-4576-9337}\and
    L.~Baldini\inst{\ref{IS4},\ref{IS5}} \orcidlink{0000-0002-9785-7726}\and
    W.~H.~Baumgartner\inst{\ref{I16}} \orcidlink{0000-0002-5106-0463}\and
    R.~Bellazzini\inst{\ref{IS4}} \orcidlink{0000-0002-2469-7063}\and
    S.~D.~Bongiorno\inst{\ref{I16}} \orcidlink{0000-0002-0901-2097}\and
    R.~Bonino\inst{\ref{IS6},\ref{IS7}} \orcidlink{0000-0002-4264-1215}\and
    A.~Brez\inst{\ref{IS4}} \orcidlink{0000-0002-9460-1821}\and
    N.~Bucciantini\inst{\ref{IS8},\ref{IS9},\ref{IS10}} \orcidlink{0000-0002-8848-1392}\and
    S.~Castellano\inst{\ref{IS4}} \orcidlink{0000-0003-1111-4292}\and
    E.~Cavazzuti\inst{\ref{I18}} \orcidlink{0000-0001-7150-9638}\and
    C.~Chen\inst{\ref{IS11}} \orcidlink{0000-0002-4945-5079}\and
    S.~Ciprini\inst{\ref{I22},\ref{IS3}} \orcidlink{0000-0002-0712-2479}\and
    E.~Costa\inst{\ref{I8}} \orcidlink{0000-0003-4925-8523}
    E.~Del Monte\inst{\ref{I8}}\orcidlink{0000-0002-3013-6334}\and
    L.~Di Gesu\inst{\ref{I18}}\orcidlink{0000-0002-5614-5028}\and
    N.~Di Lalla\inst{\ref{IS12}}\orcidlink{0000-0002-7574-1298}\and
    A.~Di Marco\inst{\ref{I8}}\orcidlink{0000-0003-0331-3259}\and
    I.~Donnarumma\inst{\ref{I18}}\orcidlink{0000-0002-4700-4549}\and
    V.~Doroshenko\inst{\ref{IS13}}\orcidlink{0000-0001-8162-1105}\and
    S.~R.~Ehlert\inst{\ref{I16}}\orcidlink{0000-0003-4420-2838}\and
    T.~Enoto\inst{\ref{IS14}}\orcidlink{0000-0003-1244-3100}\and
    Y.~Evangelista\inst{\ref{I8}}\orcidlink{0000-0001-6096-6710}\and
    S.~Fabiani\inst{\ref{I8}}\orcidlink{0000-0003-1533-0283}
    R.~Ferrazzoli\inst{\ref{I8}}\orcidlink{0000-0003-1074-8605}\and
    K.~Hayashida\inst{\ref{IS15}}\and
    J.~Heyl\inst{\ref{IS16}}\orcidlink{0000-0001-9739-367X}\and
    W.~Iwakiri\inst{\ref{IS17}}\orcidlink{0000-0002-0207-9010}\and
    S.~G.~Jorstad\inst{\ref{IS18},\ref{IS19}}\orcidlink{0000-0001-9522-5453}\and
    V.~Karas\inst{\ref{I2}}\orcidlink{0000-0002-5760-0459}\and
    F.~Kislat\inst{\ref{IC}}\orcidlink{0000-0001-7477-0380}\and
    T.~Kitaguchi\inst{\ref{IS14}}\and
    J.~J.~Kolodziejczak\inst{\ref{I16}}\orcidlink{0000-0002-0110-6136}\and
    F.~La Monaca\inst{\ref{I8}}\orcidlink{0000-0001-8916-4156}\and
    L.~Latronico\inst{\ref{IS6}}\orcidlink{0000-0002-0984-1856}\and
    I.~Liodakis\inst{\ref{I16}}\orcidlink{0000-0001-9200-4006}\and
    S.~Maldera\inst{\ref{IS6}}\orcidlink{0000-0002-0698-4421}\and
    A.~Manfreda\inst{\ref{IS21}}\orcidlink{0000-0002-0998-4953}\and  
    F.~Marin\inst{\ref{I10}}\orcidlink{0000-0003-4952-0835}\and
    A.~P.~Marscher\inst{\ref{IS18}}\orcidlink{0000-0001-7396-3332}\and
    H.~L.~Marshall\inst{\ref{IS23}}\orcidlink{0000-0002-6492-1293}\and
    F.~Massaro\inst{\ref{IS6},\ref{IS7}}\orcidlink{0000-0002-1704-9850}\and
    I.~Mitsuishi\inst{\ref{IS24}}\and
    T.~Mizuno\inst{\ref{IS25}}\orcidlink{0000-0001-7263-0296}\and
    M.~Negro\inst{\ref{IS26}}\orcidlink{0000-0002-6548-5622}\and
    C.~Y.~Ng\inst{\ref{IS27}}\orcidlink{0000-0002-5847-2612}\and
    S.~L.~O'Dell\inst{\ref{I16}}\orcidlink{0000-0002-1868-8056}\and
    N.~Omodei\inst{\ref{IS12}}\orcidlink{0000-0002-5448-7577}\and
    C.~Oppedisano\inst{\ref{IS6}}\orcidlink{0000-0001-6194-4601}\and
    A.~Papitto\inst{\ref{IS2}}\orcidlink{0000-0001-6289-7413}\and
    G.~G.~Pavlov\inst{\ref{IS28}}\orcidlink{0000-0002-7481-5259}\and
    A.~L.~Peirson\inst{\ref{IS12}}\orcidlink{0000-0001-6292-1911}\and
    M.~Perri\inst{\ref{IS3},\ref{IS2}}\orcidlink{0000-0003-3613-4409}\and
    M.~Pesce-Rollins\inst{\ref{IS4}}\orcidlink{0000-0003-1790-8018}\and
    M.~Pilia\inst{\ref{I19}}\orcidlink{0000-0001-7397-8091}\and
    A.~Possenti\inst{\ref{I19}}\orcidlink{0000-0001-5902-3731}\and
    S.~Puccetti\inst{\ref{IS3}}\orcidlink{0000-0002-2734-7835}\and
    B.~D.~Ramsey\inst{\ref{I16}}\orcidlink{0000-0003-1548-1524}\and
    J.~Rankin\inst{\ref{I8}}\orcidlink{0000-0002-9774-0560}\and
    O.~J.~Roberts\inst{\ref{IS11}}\orcidlink{0000-0002-7150-9061}\and
    R.~W.~Romani\inst{\ref{IS12}}\orcidlink{0000-0001-6711-3286}\and
    C.~Sgr\`{o}\inst{\ref{IS4}}\orcidlink{0000-0001-5676-6214}\and
    P.~Slane\inst{\ref{I6}}\orcidlink{0000-0002-6986-6756}\and
    G.~Spandre\inst{\ref{IS4}}\orcidlink{0000-0003-0802-3453}\and
    D.~A.~Swartz\inst{\ref{IS11}}\orcidlink{0000-0002-2954-4461}\and
    T.~Tamagawa\inst{\ref{IS14}}\orcidlink{0000-0002-8801-6263}\and
    F.~Tavecchio\inst{\ref{IS29}}\orcidlink{0000-0003-0256-0995}\and
    Y.~Tawara\inst{\ref{IS24}}\and
    A.~F.~Tennant\inst{\ref{I16}}\orcidlink{0000-0002-9443-6774}\and
    N.~E.~Thomas\inst{\ref{I16}}\orcidlink{0000-0003-0411-4606}\and
    A.~Trois\inst{\ref{I19}}\orcidlink{0000-0002-3180-6002}\and
    S.~S.~Tsygankov\inst{\ref{I11}}\orcidlink{0000-0002-9679-0793}\and
    R.~Turolla\inst{\ref{I12},\ref{I24}}\orcidlink{0000-0003-3977-8760}\and
    J.~Vink\inst{\ref{IS30}}\orcidlink{0000-0002-4708-4219}\and
    M.~C.~Weisskopf\inst{\ref{I16}}\orcidlink{0000-0002-5270-4240}\and
    K.~Wu\inst{\ref{I24}}\orcidlink{0000-0002-7568-8765}\and
    F.~Xie\inst{\ref{IS31},\ref{I8}}\orcidlink{0000-0002-0105-5826}
}

\institute{
    Dipartimento di Matematica e Fisica, Università degli Studi Roma Tre, Via della Vasca Navale 84, 00146 Roma, Italy \label{I1} \and 
    Astronomical Institute of the Czech Academy of Sciences, Bo\v{c}n\'{i} II 1401/1, 14100 Praha 4, Czech Republic \label{I2} \and
    Astronomical Institute, Faculty of Mathematics and Physics, Charles University, V Holešovičkách 2, Prague 8, 180~00, Czech Republic \label{I3} \and
    Institute of Theoretical Physics, Faculty of Mathematics and Physics, Charles University, V Holešovickách 2, CZ-180 00 Praha 8, Czech Republic \label{I4}\and
    Physics Department, McDonnell Center for the Space Sciences, and Center for Quantum Leaps, Washington University in St. Louis, St. Louis, MO 63130, USA \label{I5}\and
    Center for Astrophysics, Harvard \& Smithsonian, 60 Garden St, Cambridge, MA 02138, USA \label{I6}\and
    Physics Dept., CB 1105, Washington University, One Brookings Drive, St. Louis, MO 63130-4899 \label{I7}\and
    INAF Istituto di Astrofisica e Planetologia Spaziali, Via del Fosso del Cavaliere 100, 00133 Roma, Italy \label{I8}\and
    School of Mathematics, Statistics, and Physics, Newcastle University, Newcastle upon Tyne NE1 7RU, UK \label{I9}\and
    Universit\'{e} de Strasbourg, CNRS, Observatoire Astronomique de Strasbourg, UMR 7550, 67000 Strasbourg, France\label{I10}\and
    Department of Physics and Astronomy, 20014 University of Turku, Finland \label{I11}\and
    Dipartimento di Fisica e Astronomia, Universit\`{a} degli Studi di Padova, Via Marzolo 8, 35131 Padova, Italy \label{I12}\and
    Nordita, KTH Royal Institute of Technology and Stockholm University, Hannes Alfv\'ens v\"ag 12, SE-10691 Stockholm, Sweden \label{I13}\and
    California Institute of Technology, Pasadena, CA 91125, USA \label{I14}\and
    Space Research Institute of the Russian Academy of Sciences, Profsoyuznaya Str. 84/32, Moscow 117997, Russia\label{IA}\and
    INAF—Osservatorio di Astrofisica e Scienza dello Spazio di Bologna, Via P. Gobetti 101, I-40129 Bologna, Italy\label{IB}\and
    Yamagata University,1-4-12 Kojirakawa-machi, Yamagata-shi 990-8560, Japan \label{I15}\and
    NASA Marshall Space Flight Center, Huntsville, AL 35812, USA \label{I16}\and
    NASA Goddard Space Flight Center, Greenbelt, MD 20771, USA \label{I17}\and
    Agenzia Spaziale Italiana, Via del Politecnico snc, 00133 Roma, Italy \label{I18}\and
    INAF-Osservatorio Astronomico di Cagliari, via della Scienza 5, I-09047 Selargius (CA), Italy \label{I19}\and
    Univ. Grenoble Alpes, CNRS, IPAG, 38000 Grenoble, France \label{I20}\and
    Dipartimento di Fisica, Universit\`{a} degli Studi di Roma ``Tor Vergata'', Via della Ricerca Scientifica 1, 00133 Roma, Italy \label{I21}\and
    Istituto Nazionale di Fisica Nucleare, Sezione di Roma ``Tor Vergata'', Via della Ricerca Scientifica 1, 00133 Roma, Italy \label{I22}\and
    Department of Astronomy, University of Maryland, College Park, Maryland 20742, USA \label{I23}\and
    Mullard Space Science Laboratory, University College London, Holmbury St Mary, Dorking, Surrey RH5 6NT, UK \label{I24}\and
    Instituto de Astrof\'{i}sica de Andaluc\'{i}a -- CSIC, Glorieta de la Astronom\'{i}a s/n, 18008 Granada, Spain\label{IS1}\and
    INAF Osservatorio Astronomico di Roma, Via Frascati 33, 00040 Monte Porzio Catone (RM), Italy\label{IS2}\and
    Space Science Data Center, Agenzia Spaziale Italiana, Via del Politecnico snc, 00133 Roma, Italy\label{IS3}\and
    Istituto Nazionale di Fisica Nucleare, Sezione di Pisa, Largo B. Pontecorvo 3, 56127 Pisa, Italy\label{IS4}\and
    Dipartimento di Fisica, Universit\`{a} di Pisa, Largo B. Pontecorvo 3, 56127 Pisa, Italy\label{IS5}\and
    Istituto Nazionale di Fisica Nucleare, Sezione di Torino, Via Pietro Giuria 1, 10125 Torino, Italy\label{IS6}\and
    Dipartimento di Fisica, Universit\`{a} degli Studi di Torino, Via Pietro Giuria 1, 10125 Torino, Italy\label{IS7}\and
    INAF Osservatorio Astrofisico di Arcetri, Largo Enrico Fermi 5, 50125 Firenze, Italy\label{IS8}\and
    Dipartimento di Fisica e Astronomia, Universit\`{a} degli Studi di Firenze, Via Sansone 1, 50019 Sesto Fiorentino (FI), Italy\label{IS9}\and
    Istituto Nazionale di Fisica Nucleare, Sezione di Firenze, Via Sansone 1, 50019 Sesto Fiorentino (FI), Italy\label{IS10}\and
    Science and Technology Institute, Universities Space Research Association, Huntsville, AL 35805, USA\label{IS11}\and
    Department of Physics and Kavli Institute for Particle Astrophysics and Cosmology, Stanford University, Stanford, California 94305, USA\label{IS12}\and
    Institut f\"{u}r Astronomie und Astrophysik, Universität Tübingen, Sand 1, 72076 T\"{u}bingen, Germany\label{IS13}\and
    RIKEN Cluster for Pioneering Research, 2-1 Hirosawa, Wako, Saitama 351-0198, Japan\label{IS14}\and
    Osaka University, 1-1 Yamadaoka, Suita, Osaka 565-0871, Japan\label{IS15}\and
    University of British Columbia, Vancouver, BC V6T 1Z4, Canada\label{IS16}\and
    International Center for Hadron Astrophysics, Chiba University, Chiba 263-8522, Japan\label{IS17}\and
    Institute for Astrophysical Research, Boston University, 725 Commonwealth Avenue, Boston, MA 02215, USA\label{IS18}\and
    Department of Astrophysics, St. Petersburg State University, Universitetsky pr. 28, Petrodvoretz, 198504 St. Petersburg, Russia\label{IS19}\and
    Department of Physics and Astronomy and Space Science Center, University of New Hampshire, Durham, NH 03824, USA\label{IC}\and
    Istituto Nazionale di Fisica Nucleare, Sezione di Napoli, Strada Comunale Cinthia, 80126 Napoli, Italy\label{IS21}\and
    MIT Kavli Institute for Astrophysics and Space Research, Massachusetts Institute of Technology, 77 Massachusetts Avenue, Cambridge, MA 02139, USA\label{IS23}\and
    Graduate School of Science, Division of Particle and Astrophysical Science, Nagoya University, Furo-cho, Chikusa-ku, Nagoya, Aichi 464-8602, Japan\label{IS24}\and
    Hiroshima Astrophysical Science Center, Hiroshima University, 1-3-1 Kagamiyama, Higashi-Hiroshima, Hiroshima 739-8526, Japan\label{IS25}\and
    Department of Physics and Astronomy, Louisiana State University, Baton Rouge, LA 70803, USA\label{IS26}\and
    Department of Physics, University of Hong Kong, Pokfulam, Hong Kong\label{IS27}\and
    Department of Astronomy and Astrophysics, Pennsylvania State University, University Park, PA 16801, USA\label{IS28}\and
    INAF Osservatorio Astronomico di Brera, via E. Bianchi 46, 23807 Merate (LC), Italy\label{IS29}\and
    Anton Pannekoek Institute for Astronomy \& GRAPPA, University of Amsterdam, Science Park 904, 1098 XH Amsterdam, The Netherlands\label{IS30}\and
    Guangxi Key Laboratory for Relativistic Astrophysics, School of Physical Science and Technology, Guangxi University, Nanning 530004, China\label{IS31}
}


 
\abstract{
We present the results of the first X-ray polarimetric observation of the low-mass X-ray binary \mbox{4U 1957+115}, performed with the Imaging X-ray Polarimetry Explorer in May 2023. The binary system has been in a high-soft 
spectral state since its discovery and is thought to host a black hole. The $\sim$571~ks observation reveals a linear polarisation degree of $1.9\% \pm 0.6$\% and a polarisation angle of $-41\fdg8 \pm 7\fdg9$
in the 2--8 keV energy range. Spectral modelling is consistent with the dominant contribution coming from the standard accretion disc, while polarimetric data suggest a significant role of returning radiation: photons that are bent by strong gravity effects and forced to return to the disc surface, where they can be reflected before eventually reaching the observer 
. In this setting, we find that models with a black hole spin lower than 0.96 and an inclination lower than $50\degr$ are disfavoured.}


   \maketitle
%

\section{Introduction}

Low-mass X-ray binaries are binary star systems consisting of a compact object, such as a black hole (BH) or a neutron star, and a low-mass companion from which mass is transferred via Roche-lobe overflow. Most of these systems typically display strong variability in their X-ray emission. 
A notable exception to this behaviour is represented by 4U~1957+115. Discovered in 1973 by the \emph{Uhuru} satellite during its scan of the Aquila region \citep{Giacconi+74}, the source is exceptional for being one of a few historically persistently active BH candidates. This short list also includes \mbox{LMC X-1}, \mbox{LMC X-3}, \mbox{Cyg X-1}, and \mbox{Cyg X-3}, which are, unlike \mbox{4U 1957+115}, classified as high-mass X-ray binary systems \citep{Orosz+09,Orosz+14,Miller-Jones+21}. In line with the first two of those sources, \mbox{4U 1957+115} is always in 
a soft X-ray spectral state \citep[e.g.][]{Yaqoob+93,Ricci+95,Nowak+99,Nowak+08,Nowak+12,Maitra+14,Sharma+21,Barillier+23}.
Furthermore, the source has never shown any observable radio jet \citep{Russell+11}, which aligns with the source's persistent soft state behaviour. Remarkably, the absence of any radio hot spot detection has led to the establishment of the most rigorous upper limits on the radio-to-X-ray flux ratio for a BH in a soft state \citep{Maccarone+20}.

Unfortunately, limited information is available regarding the system's mass, distance, and inclination; this is due to its persistent nature, which hampers optical measurements of binary parameters, best done during quiescence. Optical emission is likely dominated by the accretion disc \citep{Hakala+14}, but optical observations by \citet{Thorstensen+87} revealed a nearly sinusoidal orbital variation with a period of $9.329 \pm 0.011$ hr and $\pm 20$\% orbital modulation. 
Several lines of argument have attributed this phenomenon to the irradiation of the companion star's surface \citep{Margon+78} by the accretion disc of the compact object \citep{Bayless+11,Mason+12,Gomez+15}. This suggests that, from the perspective of the primary, the secondary star occupies a substantial solid angle, which implies a relatively small separation and thus a relatively low total mass for the binary system. This is consistent with the primary being either a neutron star  \citep{Bayless+11} or a low-mass BH \citep[$M < 6.2 \ \Msun$;][]{Gomez+15}; both possibilities are allowed by the $\approx$ 0.25--0.3 mass ratio derived by \citet{Longa-pena+15} through Bowen fluorescence line studies. Although no study of the nature of the compact object has been conclusive, the lack of Type I bursts, pulsations, `surface emission' components, or signatures of a boundary layer emission in the X-ray spectra of the source disfavour the neutron star hypothesis \citep{Maccarone+20}. 

According to interstellar medium absorption modelling along the line of sight, the source is believed to be located outside the Galactic plane, at a minimum distance of approximately 5~kpc \citep{Nowak+08,Yao+08}. In a recent study by \citet{Barillier+23}, who analysed the parallax and proper motion data from the \textit{Gaia} Early Data Release 3 catalogue \citep{GaiaColl+21} assuming that 4U~1957+115 is located in the Galactic halo, they found that the distance probability distribution peaks at 6~kpc. However, the study revealed a substantial cumulative probability (17\%) in the range of 15 to 30~kpc. From this, and from {\it NuSTAR} 
data analysis, they derived a mass probability distribution for the source, 50\% of which corresponds to $M<7.2 \Msun$. Notably, a significant portion of the mass probability distribution (22\%) is found to lie within the `mass gap' of 2--5\,$\Msun$, which is known to contain only a limited number of sources \citep{Ozel+10,Farr+11,Gomez+15}. 

The absence of eclipses and any orbital modulations in the X-ray light curve \citep{Wijnands+02} allowed for the estimate of an upper limit of between $65\degr$ and $\approx$75\degr\ for the source inclination, which is consistent with the model of optical variability \citep{Hakala+99}. Furthermore, this inclination range is also in agreement with the absence of a highly ionised wind \citep{Ponti+12,Parra+23}. On the other hand, X-ray spectral fitting analyses tend to predict large values for the system's inclination, such as $\sim$78\degr
\ \citep{Maitra+14}. Conversely, several optical modulation studies favour systems with lower inclinations \citep[e.g. $\sim$13\degr; 
][]{Gomez+15}. 

As a soft-state source, 4U~1957+115's X-ray spectrum is dominated by the accretion disc emission, with a minor contribution from a Comptonisation component and weak reflection features \citep{Sharma+21}. A correlation has been observed between the brightness of the source and the contribution of the Comptonised component; by analysing  {\it NuSTAR} observations, \citet{Barillier+23} described the increasing 
behaviour of the hard tail with rising flux with two different tracks, with one having significantly stronger tails than the other \citep[e.g. see Figs. 8--10 in][]{Barillier+23}. Their proposed explanation for this correlation is a reduction in the disc hardening factor associated with the increase in the amplitude of the power-law tail; this scenario suggests that electron scattering in a hot corona becomes more important as it diminishes in the upper layers of the optically thick accretion disc. Through the analysis of the reflection component and continuum fitting of the disc component, several estimates of the BH spin in 4U~1957+115 have been obtained. These estimates consistently describe the source as rapidly rotating, with spin values as high as $a>0.9$ \citep{Nowak+12}, $a>0.98$ \citep{Maitra+14}, $a\sim 0.85$ \citep{Sharma+21}, and $a=0.95$ \citep{Draghis+23}. 

An additional way to obtain valuable information about sources of this class is offered by X-ray polarimetry, which is sensitive to the geometry of the emitting region. In the case of soft-state sources, several studies have proposed the potential use of polarimetric data to constrain the inclination of the accretion disc and the  BH spin \citep{Connors+77,Stark+77,Connors+80,Dovciak+04,Dovciak+08,Schnittman+09,Schnittman+10,Taverna+20,Taverna+21,KrawczynskiBeheshtipour+22,Mikusincova+23,Loktev2023}. The Imaging X-ray Polarimetry Explorer \citep[IXPE;][]{Weisskopf+22}, a space-based observatory launched on 2021 December 9, has enabled, for the first time, a sensitive X-ray polarimetric study of this class of source. \ixpe\ has already observed several BH X-ray binaries, including Cyg X-1 in the hard state \citep{Krawczynsky+22}, Cyg X-3 in the hard and intermediate states with strong reflection features \citep{Veledina+23}, LMC X-1 in the soft state  \citep{Podgorny+23}, and 4U 1630$-$47 in both the high-soft
\citep{Ratheesh+23} and the very high-steep 
power-law state \citep{RodriguezCavero+23}.

We present the first X-ray polarimetric measurement of \mbox{4U 1957+115}, which was observed by \ixpe in May 2023. Simultaneous X-ray observations were also carried out with NICER 
\citep{Gendreau+12}, {\it NuSTAR} \citep{Harrison+13}, and SRG/ART-XC 
\citep{Pavlinsky+21}, which provided a better spectral coverage. The paper is organised as follows. Section~\ref{sec:DataRed} describes the observations and the data reduction techniques. Our spectral and polarimetric analysis is presented in Sect.~\ref{sec:DataAnalysis}, while Sect.~\ref{sec:SpinIncCost} describes the modelling of the polarimetric data assuming different BH spin and system inclination values. In Sect.~\ref{sec:kerrC} we analyse the impact of corona emission on the observed polarisation properties of the source. Finally, Sect.~\ref{sec:Conclusions} summarises the main results of our analysis.

\section{Observation and data reduction}
\label{sec:DataRed}

\mbox{4U 1957+115} was observed by \ixpe from 2023 May 12--24 (ObsID: 02006601) for a net exposure time of $\sim$571~ks. \ixpe detectors \citep{Soffitta+21} can measure the Stokes parameters $I$, $Q$ and $U$, and they have imaging capabilities that allow for the spatial separation of source and background regions. We obtained Level 2 event files from the HEASARC archive\footnote{Available at \url{https://heasarc.gsfc.nasa.gov/docs/ixpe/archive/}.} and subsequently filtered them for source and background regions using the \texttt{xpselect} tool from the \textsc{ixpeobssim} software package \citep[version 30.5,][]{Baldini+22}. For the source extraction regions, circular regions with a radius of 60\arcsec\ were chosen for each detector unit. The background regions were defined as annuli with an inner radius of 180\arcsec\ and an outer radius of 280\arcsec. We utilised the \textsc{ixpeobssim} \texttt{PHA1} algorithm to generate weighted Stokes $I$, $Q$, and $U$ parameters, which were binned into 11 bins with sizes 0.5 keV in 2--7 keV, except for the last 7--8 keV bin, for which it was 1 keV 
Polarisation cubes were generated using the unweighted \texttt{pcube} algorithm \citep{Baldini+22}. In this way, the energy-dependent polarisation degree (PD) and polarisation angle (PA) were created. To ensure significant detection in all bins except the last one, we created the PD and PA for five energy bins with a bin size of 1 keV in 2--6 keV for all but the last, 6--8 keV bin, which was 2 keV

NICER is a large area 0.2--12 keV X-ray timing mission on the International Space Station.  Its X-ray Timing Instrument (XTI) is composed of 56 co-aligned focal-plane modules (FPMs), 52 of which have been functional since its launch in 2017.  Each FPM houses a silicon drift detector and is paired to a single-bounce concentrator optic. The XTI is collimated to sample a field of view approximately 3~arcmin in radius.  Because NICER is a non-imaging instrument, the background is modelled rather than sampled directly \citep{Remillard2022}.  We adopted the SCORPEON background model in our analysis,\footnote{\url{https://heasarc.gsfc.nasa.gov/docs/nicer/analysis_threads/scorpeon-xspec/}} via the \texttt{niscorpspect} utility.

NICER observed 4U~1957+115 over the duration of the \ixpe campaign, in continuous observations typically lasting $\sim 10$~min, up to 40 min.  These observations were processed using \texttt{nicerl2} with standard screening except for the undershoot and overshoot rate filters, which were left unrestricted during this initial stage of the processing.  Data during South Atlantic Anomaly passages were reduced separately, but not automatically excluded from analysis.  All resulting observations were separated into continuous interval good time intervals (GTIs).  For each GTI, the per-FPM distributions of undershoot, overshoot, and X-ray rates were compared across the detector ensemble and any detector presenting a $>$10 robust standard deviation outlier for any of the rates was excised from analysis.  Between one and seven detectors were screened out for each interval, owing to elevated undershoot rates associated with optical contamination.  Detector 63 was particularly affected.  
Events from the remaining detectors were summed to produce spectra with associated response products.  Any GTI of $<100$~s duration or exhibiting an elevated background (as screened by eye) was removed.  Surviving the screening, in total we obtain approximately 58~ks of good NICER time, 96 GTIs, spread among 12 ObsIDs.  Spectra and responses of all GTIs for a given ObsID were summed in weighted combination for analysis.

{\it NuSTAR} \citep{Harrison+13} observed the source with its two co-aligned X-ray telescopes with corresponding Focal Plane Module A (FPMA) and B (FPMB) in three separate observations. The net exposure times for these observations were $18.7$ ks (ObsID: 30902042002), $20.2$ ks (ObsID: 30902042004) and $19.7$ ks (ObsID: 30902042006), respectively. We generated cleaned event files using the dedicated \texttt{nupipeline} task and the most recent calibration files (CALDB 20230516). We performed background subtraction extracting the background from a circular region with a standard radius of 60\arcsec\ for both detectors and for all observations. The source extraction radius was set at 123\arcsec, 123\arcsec\ and 118\arcsec\ for the three observations following a procedure that maximises the signal-to-noise ratio \citep{Piconcelli+04}. Subsequently, we re-binned the spectra using the standard task \texttt{ftgrouppha}, implementing the optimal scheme proposed by \citet{Kaastra+16}, with the additional requirement of a minimum signal-to-noise ratio of 3 in each bin. The FPMA and FPMB spectra were fitted independently in the spectral analysis, but are combined in plots for clarity. 

The \textit{Mikhail Pavlinsky }ART-XC telescope \citep{Pavlinsky+21} on board the SRG observatory \citep{Sunyaev+21} observed the source twice on May 13 and May 21, for 68 and 67 ks, respectively. The data were reduced with the {\sc artproducts} v1.0 package and CALDB version {\sc v20220908}. 
Light curves in 4--8 and 8--16 keV bands were extracted from a 2 arcmin  circular region, centred on the source. Data from all seven mirror modules were combined.

Throughout the entire paper, uncertainties are reported at the 
$90$\% confidence level, unless explicitly mentioned otherwise. Upper and lower limits are provided at the $99.7$\% ($3 \sigma$) confidence level for one parameter of interest.

\section{Data analysis}\label{sec:DataAnalysis}
\subsection{Timing and spectral analysis} \label{Sec:SpectralAnalysis}

\begin{figure}
\centering
\includegraphics[width=\columnwidth]{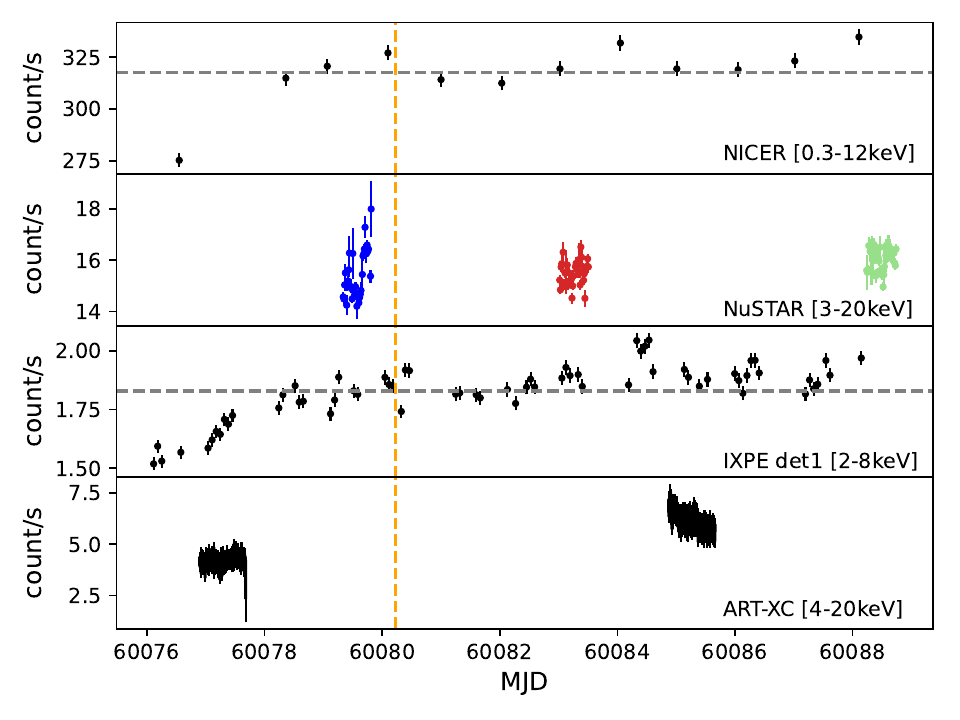}
\caption{Light curves of 4U~1957+115 as seen by {NICER} in the 0.3--12 keV energy band, {\it NuSTAR} in 3--20 keV, \ixpe in 2--8 keV, and {ART-XC} in  4--12 keV. {\it NuSTAR} data  coloured in blue, red, and green refer to the three epochs described in the spectral analysis and used in the spectra shown in Fig. \ref{fig:Spectra}. The vertical orange line shows the subdivision of the \ixpe observation  described in Sect.~\ref{sec:Pol_Data}, and the dashed horizontal lines in the {NICER} and \ixpe light curves indicate the mean values of the count rate.}
    \label{fig:lightcurves}
\end{figure}

Simultaneously with \ixpe observations, we monitored the source in the X-rays with the following instruments and energy bands:  NICER (0.3--12 keV), {\it NuSTAR} (3--20 keV) and SRG/ART-XC (4--30 keV). \ixpe and NICER cover the soft X-ray band and {\it NuSTAR} and {ART-XC} the hard X-ray band over a period of 14 days. The binning size of each instrument is 1~ks for {\it NuSTAR}, 623 s for {NICER}, and 6~ks for \ixpe. We see from Fig.~\ref{fig:lightcurves} that the flux significantly increases in the soft X-ray band on the first three days of monitoring with {NICER} and \ixpe, and then fluctuates around an average value on the last 10 days. In the hard X-ray band, the count rate for {\it NuSTAR} appears generally constant within the error bars, while a count rate increase is observed between the two {ART-XC} observations.

\begin{figure}
    \centering
    \includegraphics[width=\columnwidth]{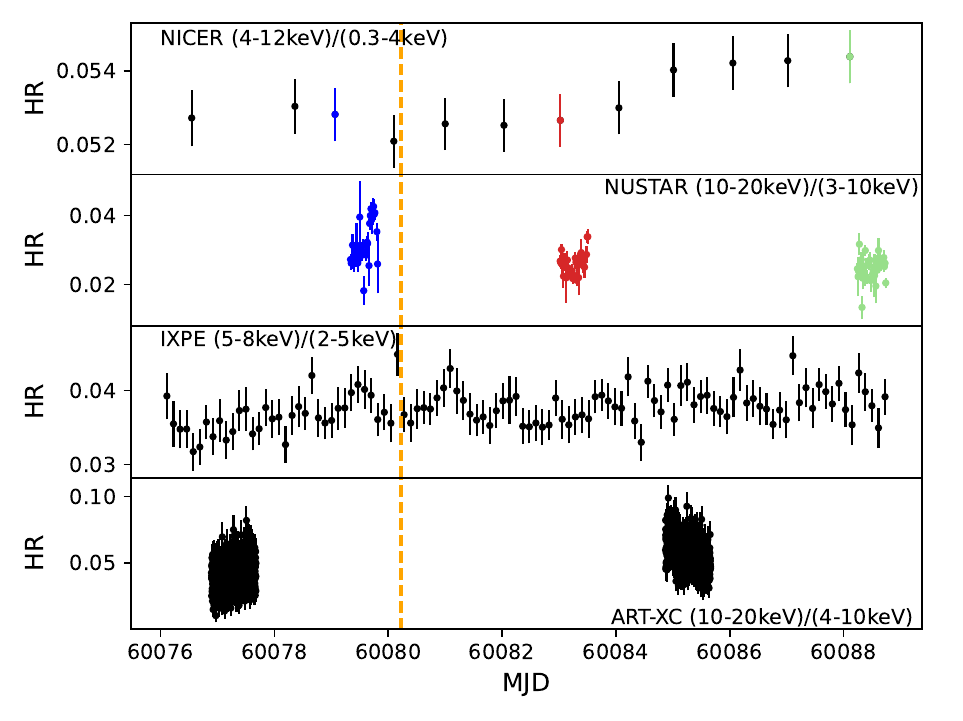}
    \caption{Time evolution of the hardness ratio from {NICER}, {\it NuSTAR}, \ixpe, and {ART-XC} data as defined in the text. {\it NuSTAR} and NICER data  coloured in blue, red, and green refer to the three epochs described in the spectral analysis and used in the spectra shown in Fig. \ref{fig:Spectra}. The vertical orange line shows the subdivision of the \ixpe observation described in Sect.~\ref{sec:Pol_Data}.}
    \label{fig:HR}
\end{figure}

In order to analyse the change of state of the source, we calculate the hardness ratio defined as the ratio between the hard energy band over the soft energy band. We define the respective hard/soft energy bands for each instrument: \ixpe 5--8 keV/2--5 keV, {NICER} 4--12 keV/0.3--4 keV, {\it NuSTAR} 10--20 keV/3--10 keV, and {ART-XC} 10--20 keV/4--10 keV. Figure \ref{fig:HR} shows the evolution of the hardness ratio calculated from the \ixpe, {NICER}, {\it NuSTAR,} and {ART-XC} data. The \ixpe hardness ratio fluctuates between 0.030 and 0.045 over the whole period of observation. For {NICER}, the hardness ratio varies between 0.052 and 0.054. By calculating the null hypothesis probability fitted with a constant, we have a p-value of 0.0024 for NICER and $\mathrm{4.7003 \cdot 10^{-5}}$ for IXPE. Therefore, for NICER the hardness ratio remains constant within the error bars but is more variable for \ixpe. Regarding {\it NuSTAR}, the hardness ratio decreases from 0.045 to 0.02 in 10 days, indicating a slight transition of the source towards a softer state. The {ART-XC} hardness ratio slowly increases at the start of the observations and decreases at the end, which is in agreement with the results from {\it NuSTAR}. 

For the spectral analysis of our source, we focused on {\it NuSTAR} and NICER simultaneous observations, as indicated in Figs.~\ref{fig:lightcurves} and \ref{fig:HR}, performing a joint fit of the spectra of the two satellites. As the {\it NuSTAR} high energy flux decreases during the three observations, we adopted different energy ranges according to the energy at which the background starts dominating: 3--30~keV in Period 1, 3--25~keV in Period 2 and 3--20~keV in Period~3. For the NICER data, we fitted the spectra in the 0.7--8~keV energy range for all periods. We used the \textsc{xspec} package \citep[v12.13.0c;][]{Arnaud+96} and employed the following model in the analysis:
\begin{equation}
\tt tbabs*(kerrbb+expabs*powerlaw).
    \label{eq:SpectralFit}
\end{equation}

In addition, a cross-calibration constant was included to account for discrepancies between the {\it NuSTAR} FPMA, FPMB, and NICER spectra. This constant was kept fixed at 1 for the {\it NuSTAR} FPMA, while the best-fitting values for {\it NuSTAR} FPMB and NICER are 
$0.981 \pm 0.005$ and $0.940 \pm 0.005$, respectively. The spectral model includes a {\tt tbabs} \citep{Wilms+00} component to account for interstellar absorption. A {\tt kerrbb} component is used to describe the accretion disc emission, properly accounting for relativistic effects \citep{Li+05}; this component describes the radiation emitted from the accretion disc assuming that its structure and its emission properties are described accurately by the standard \citet{Novikov1973} model. A {\tt powerlaw} component was included as a phenomenological representation of the Comptonised emission originating from the corona, which was convolved with an {\tt expabs} component to include a low-energy roll-off. The roll-off energy was obtained in a preliminary analysis using {\tt diskbb} in place of {\tt kerrbb} and equating it to the inner disc temperature. This initial modelling with {\tt diskbb} revealed a large disc emission peak temperature ($1.39 \pm 0.02$, $1.41 \pm 0.02,$ and $1.44 \pm 0.02$ keV in Periods 1, 2, and 3, respectively), which are typical for this source \citep{Sharma+21, Barillier+23}.

\begin{table*}
\begin{center}
\begin{tabular}[t]{cccccc}
\hline
\hline
 Component & Parameter (unit) & Description & & Values &\\
\cline{4-6}
 & & & Period 1 & Period 2 & Period 3 \\

\hline
 
{\tt tbabs} & $N_{\rm H} (10^{22} \ \mathrm{cm}^{-2})$ & Hydrogen column density & $0.17 \pm 0.01$ & $0.18 \pm 0.01$ & $0.19 \pm 0.01$ \\
\hline

{\tt kerrbb} & $\eta$ & Inner-torque modification & $0^\dagger$ & - & -\\
& $a$ & BH spin & $0.992 \pm 0.003$ & - & -\\
& $i$  (deg) & Inclination & $75^\dagger$ & - & - \\
& $M_{\rm BH} \ (\mathrm{M_\odot)}$ & BH mass & $4.6^\dagger$ & - & -\\

& $\dot{M} \ (10^{16} \mathrm{g\,s^{-1}})$ &  Mass accretion rate & $3.47 \pm 0.04$ & $3.59 \pm 0.04$ & $3.72 \pm 0.05$ \\

& $D$  (kpc) & Distance & $7.8^\dagger$ & - & -\\
& $hd$ & hardening factor & $1.7^\dagger$ & - & -\\
& $r_{\rm flag}$ & Self-irradiation & $1^\dagger$ & - & - \\
& $s_{\rm flag}$ & Limb-darkening & $0^\dagger$ & - & -\\
& norm & Normalisation & $1^\dagger$ & - & -\\
\hline

  
{\tt expabs} & $E_{\rm C}$ (keV) & e-folding energy & $1.39 \pm 0.02$ & $1.41 \pm 0.02$ & $1.44 \pm 0.02$ \\
\hline

{\tt powerlaw} & $\Gamma$ & Photon index & $1.93 \pm 0.21$ & - & -\\
& norm ($10^{-3}$) & Normalisation & $10.05 \pm 4.67$ & $4.74 \pm 2.11$ & $2.79 \pm 1.29$ \\
\hline
& $\chi^2$/dof & & & 975.4/951 \\
\hline

\end{tabular}
\end{center}
\caption{Best-fitting parameters obtained in the spectral analysis of NICER and {\it NuSTAR} data during the three periods of observation. The {\tt expabs} e-folding energy was obtained from an initial modelling using {\tt diskbb}. Parameters marked with $^\dagger$ are kept frozen in the spectral analysis.}
\label{Tab:SpectralFit}
\end{table*}

Disc continuum fitting often encounters substantial degeneracy among various spectral parameters. These parameters encompass the BH mass, distance, accretion rate, hardening factor, system inclination and BH spin. This challenge is notably pronounced in the case of this source, primarily because of the limited availability of robust constraints regarding mass and distance \citep{Barillier+23}. 
Several analyses disfavour configurations with low spin and/or inclination values due to the broad spectral peak in the disc emission typically observed in this source \citep{Maitra+14, Sharma+21}. In our analysis, we initially left the system inclination free to vary in the fitting procedure; the best fit was obtained for the maximum value allowed by the model ($i= 85\degr$). However, the lack of any X-ray evidence for binary orbital modulation \citep{Wijnands+02} suggests that the source inclination cannot exceed $\approx 75\degr$; thus we decided to freeze the inclination of the system to $i= 75\degr$, following the approach used in the X-ray analysis performed by \citet{Nowak+08,Nowak+12}. We kept the BH spin, its mass and the distance free to vary in the fitting procedure, together with the accretion rate, while we assumed a value of $1.7$ for the hardening factor. Due to the strong degeneracy between mass and distance, however, this procedure yielded very large uncertainties on both parameters.
Since the main purpose of this work is to analyse the polarimetric data of our source, for the sake of simplicity we decided to fix the mass and distance to the best fiducial values obtained by \cite{Barillier+23} combining \textit{Gaia} parallax measurements with the {\it NuSTAR} spectral analysis: $M_{\rm BH}= 4.6 \ \mathrm{M_\odot}$ and $D=7.8$ kpc. Additionally, due to the decline in high-energy flux during the second and third {\it NuSTAR} observations (see Fig.~\ref{fig:HR}), the {\tt powerlaw} photon index $\Gamma$ became difficult to constrain in these periods. Hence, we linked it across all three observations, while permitting the {\tt powerlaw} normalisation to vary independently for each period.

\begin{figure}
    \centering
    \includegraphics[width=\columnwidth]{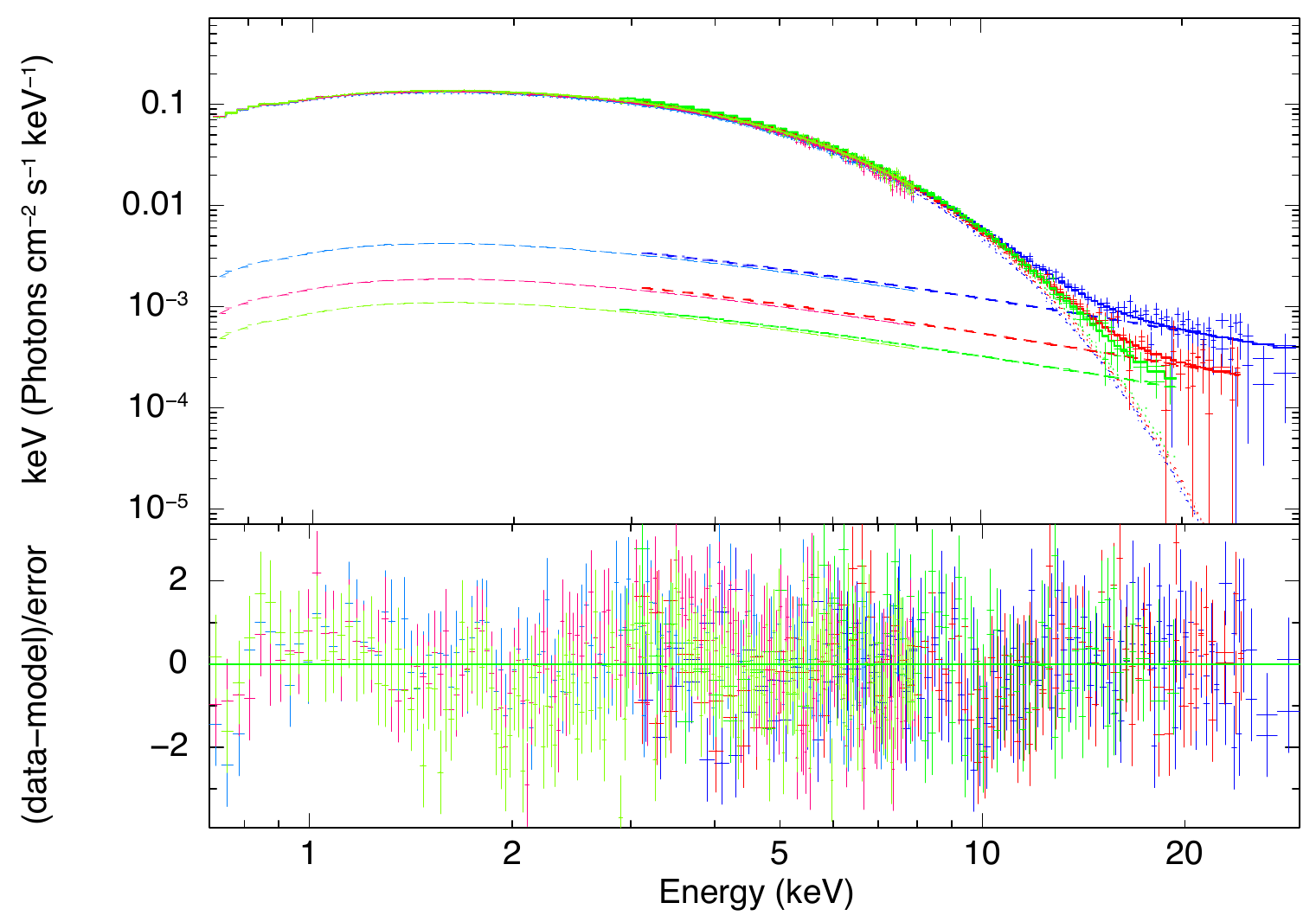}
    \caption{NICER and {\it NuSTAR} spectra of 4U~1957+115. \emph{Top panel}: Unfolded spectra (i.e. the flux $F(E)$) for the best-fitting model described by Model~\ref{eq:SpectralFit} during Periods 1,  2, and 3, shown in red, blue, and green, respectively. The total model for each period, the contributions of the {\tt kerrbb}  and the {\tt powerlaw} models are shown with the solid, dotted and dashed lines, respectively. \emph{Bottom panel}: Model minus data residuals in units of $\sigma$.}
    \label{fig:Spectra}
\end{figure}

However, the fit is statistically unacceptable with $\chi^2$/dof=2136/953, primarily due to substantial residuals in NICER spectra below 3 keV. These residuals are usually attributed to calibration issues, as similar occurrences have been noted in past observations of this source \citep{Barillier+23} and other accreting BHs \citep{Podgorny+23, RodriguezCavero+23}. Given that Model \ref{eq:SpectralFit} effectively describes {\it NuSTAR} data ($\chi^2$/dof=470/438), we decided to address the large residuals by adjusting the response file gains in NICER spectra (using the \texttt{gain fit} command in \textsc{xspec})
\footnote{The NICER response file gain parameter values from the spectral analysis are 
$1.03 \pm 0.01$ for the slope and $(-7.38 \pm 0.26) \times 10^{-2}$ keV for the offset.}. Moreover, we assigned 1\% systematic uncertainties for the NICER datasets, within the mission recommendations,\footnote{NICER calibration recommendations can be found at \url{https://heasarc.gsfc.nasa.gov/docs/nicer/analysis threads/cal-recommend/}} resulting in a revised $\chi^2$/dof=975.4/951.
The optimal spectral parameter values are detailed in Table~\ref{Tab:SpectralFit}, and the unfolded spectra along with the data-model residuals are shown in Fig.~\ref{fig:Spectra}\footnote{Due to the strong degeneracy between the hardening factor and the BH spin parameter we further investigated if the high spin scenario depicted by the spectral fit remained consistent assuming different values for the $hd$ parameter. When setting $hd=2$ our analysis yielded a BH spin of $0.953^{+0.032}_{-0.004}$, with a $\chi^2$/dof of 979.5/951. On the other hand, when considering $hd =1.5$, we obtained a lower limit for the BH spin of $0.997$, with a $\chi^2$/dof of 973.9/951.}.

Our spectral analysis allowed us to decompose the spectra into a dominating soft component, representing the accretion disc emission, and a weak hard tail, describing the photons scattered in the corona. While the disc accretion rate exhibits only a slight variation between the three {\it NuSTAR} observations, the {\tt powerlaw} normalisation shows a decrease between the first and the second periods, as suggested by the hardness ratio shown in Fig.~\ref{fig:HR}. This is further reflected by the 2--8 keV flux contribution of the hard component, which goes from 2.3\% in Period 1 to 1.0 and 0.7\% in Periods 2 and 3, respectively. Significantly, this analysis did not reveal any discernible reflection features. However, it is noteworthy that previous studies of this source have demonstrated that incorporating relativistic reflection models often enhances the overall fit quality \citep{Draghis+23}. Nonetheless, delving into this detailed investigation lies beyond the intended scope of this paper and is deferred to a future publication.

\subsection{Polarimetric analysis}\label{sec:Pol_Data}

\begin{figure*}
    \centering  \includegraphics[width=\columnwidth]{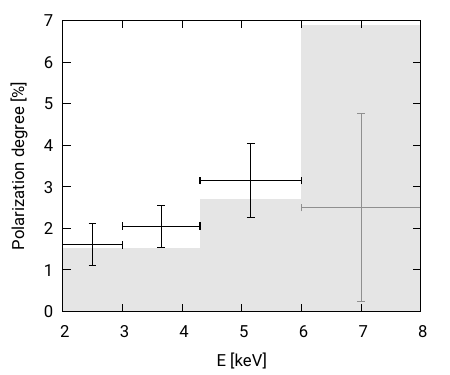}
    \includegraphics[width=\columnwidth]{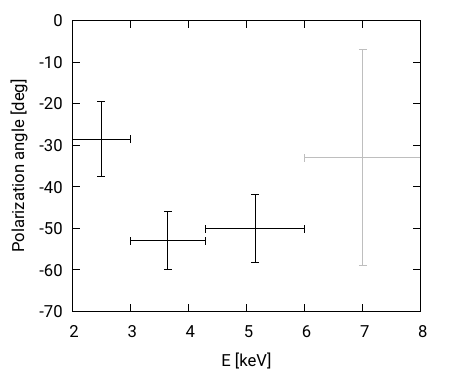}
    \caption{Measured PD (\emph{left}) and PA (\emph{right}) of 4U~1957+115, shown with $1\sigma$ error bars. The shaded grey area in the PD-plot is an estimate of the MDP$_{99}$, which shows a significant polarisation measurement from 2 keV up to 6 keV.}
    \label{fig:PD-PA-vsE}
\end{figure*}

The \ixpe observation of 4U~1957+115 revealed an average PD in the 2--8 keV band  of $1.9\% \pm 0.6$\% at a PA of $-42\fdg2 \pm 7\fdg9$, with a statistical significance of $5.2\sigma$. The measurement exceeds the minimum detectable polarisation threshold, MDP$_{99}$, which is the degree of polarisation that can be determined with a 99\% probability against the null hypothesis \citep{Weisskopf+10}. In our observation, the MDP$_{99}$ within the 2--8 keV band is 1.14\%. Figure \ref{fig:PD-PA-vsE} displays the time-averaged polarisation properties in four energy bands: 2--3, 3--4.3, 4.3--6, 6--8 keV; the first three bins show a slight increase in PD with energy, while the fourth bin shows data below the MDP$_{99}$ for that energy range, resulting in an upper PD limit. Meanwhile, the PA exhibits a decline within the 2--3 keV and the 3--4.3 keV energy bands, after which it remains relatively constant within statistical uncertainties.

Since the spectral analysis revealed a variation of the hard component during the first part of the \ixpe observation, we tried to subdivide the \ixpe data to investigate possible time variability of the polarisation. We found that in the initial part of the observation, marked in Figs.~\ref{fig:lightcurves} and \ref{fig:HR} and roughly corresponding to the increase in flux detected in \ixpe and NICER light curves, we were unable to significantly detect polarisation, as the polarisation strength was below the MDP$_{99}$ of 2.06\% in that time interval. Subsequently, throughout the remaining observation period, the polarisation properties remained steady, with a PD slightly exceeding that calculated for the entire duration of the \ixpe observation (refer to Table~\ref{Tab:PD_PA}). Because the polarisation properties identified in these distinct periods aligned, within statistical uncertainties, with the findings from the total observation, for the sake of simplicity we decided to conduct our polarimetric analysis using the entire \ixpe observation dataset. This choice is also motivated by the relatively little variation of the accretion disc emission observed in the spectral analysis, leading us to assume that the polarimetric properties of the soft component do not exhibit significant variability during the observation.

\begin{table}
\begin{center}
\begin{tabular}[t]{cccc}
\hline
\hline
Time interval & PD & PA \\
& (\%) & (deg) \\
\hline

Total  & $1.9 \pm 0.6$ & $-42.2 \pm 7.9$\\
\hline

First period  & $<2.2$ & Unconstrained\\

Second period  & $2.4 \pm 0.7 $ & $-40.6 \pm 11.1$\\ 
\hline
  
\end{tabular}
\end{center}
\caption{PD and PA found in the entire \ixpe observation and when subdividing it into two periods, as marked in Figs. \ref{fig:lightcurves} and \ref{fig:HR}. }
\label{Tab:PD_PA}
\end{table}

\subsection{Spectro-polarimetric analysis}\label{sec:Pol_Fit}

\begin{figure}
    \centering
    \includegraphics[width=\columnwidth]{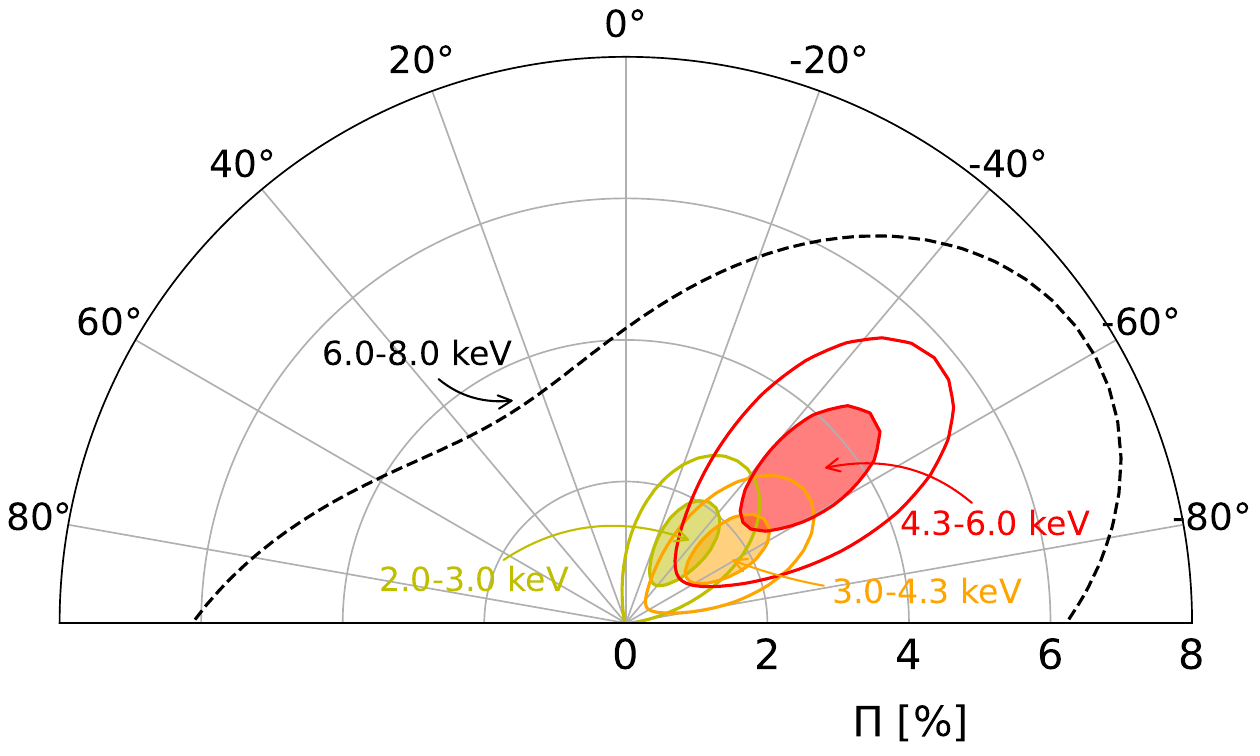}
    \caption{Polar plot of the PD and PA, assuming the spectral best-fit model, in four energy bins: 2.0--3.0, 3.0--4.3, 4.3--6.0, and 6.0--8.0~keV. The shaded and unshaded regions show the 68\% and 99.9\% confidence areas, respectively, in the first three energy bins. The dashed line indicates the 99.9\% confidence level upper limit in the fourth energy bin.}
    \label{fig:PolarPDPA}
\end{figure}

We now incorporate the polarimetric information provided by \ixpe into our spectral fit. In this section, we take the first exploratory step of fitting a spectro-polarimetric model to the \ixpe $Q$ and $U$ spectra. For this purpose, we first incorporated \ixpe $I$ spectra into the fitting procedure using Model \ref{eq:SpectralFit}. Given that the \ixpe observation extends over a longer time frame compared to the {\it NuSTAR} and NICER observations utilised in the spectral fitting detailed in Sect. \ref{Sec:SpectralAnalysis}, we made the decision to maintain all spectral parameters fixed at the values documented in Table \ref{Tab:SpectralFit}. The only exceptions to this were the disc accretion rate and the normalisation of the {\tt powerlaw} component, which we allowed to vary, together with \ixpe data cross-calibration constants. The values obtained for the  mass accretion rate and for the {\tt powerlaw} component normalisation are shown in Table \ref{Tab:Spectro-pol_FIT}; both are consistent with the values obtained in the spectral analysis described in Sect. \ref{Sec:SpectralAnalysis}. The best-fitting values of the calibration constant are $0.82 \pm 0.01$, $0.78 \pm 0.01,$ and $0.72 \pm 0.01$ for \ixpe DU1, DU2, and DU3, respectively. As was already noticed in other accreting BHs \citep{Krawczynsky+22,Podgorny+23,RodriguezCavero+23}, a simple constant is not enough to account for cross-calibration uncertainties between \ixpe, NICER, and {\it NuSTAR}; for this reason, we performed a fit on the response file gains of \ixpe spectra.\footnote{The \ixpe response file gain parameter values from the spectral analysis are: DU1 slope $0.99 \pm 0.01$, offset $(-1.52 \pm 0.25) \times 10^{-2}$ keV; DU2 slope $0.98 \pm 0.01$, offset $(2.23 \pm 0.34) \times 10^{-2}$ keV; DU3 slope $0.99 \pm 0.01$, offset $(1.46 \pm 0.21) \times 10^{-2}$ keV.} The fit resulted in a $\chi^2$/dof=436/442. As a following step, we incorporated the \ixpe $Q$ and $U$ spectra in our analysis, adopting our best-fit spectral model and applying the same gains and the same cross calibration factors as for the $I$ spectra. We assigned a constant PD and PA to the model using a {\tt polconst} component \footnote{\url{https://heasarc.gsfc.nasa.gov/xanadu/xspec/manual/node212.html}}, and performed a fit in the same four energy bands defined in Sect.~\ref{sec:Pol_Data}, leaving only the PD and PA of the {\tt polconst} model as free parameters. Figure \ref{fig:PolarPDPA} shows the contours, calculated using 50 steps for each parameter, in the polar plot of PD and PA. The PD shows a slight increase with energy as in Fig.~\ref{fig:PD-PA-vsE}, while the PA is found to have a constant behaviour, within statistical uncertainties. To determine the statistical significance of the observed increase in PD, we conducted a comparison by fitting the Q and U spectra across the entire IXPE energy range. We considered scenarios where PD and PA were either held constant or allowed to vary. When both PD and PA were held constant across energy, the resulting $\chi^2$/dof was 82.7/64. Allowing PD to vary while keeping PA constant improved the fit ($\chi^2$/dof=74.6/63), while permitting changes in both PD and PA did not significantly enhance the fit, yielding $\chi^2$/dof=72.7/62. Using an F-test to compare these models, we found that the model that allows PD to vary while maintaining a constant PA is preferred over the model with constant PD and PA, at a confidence level of $99$\%.

We then proceeded to incorporate in the fit a physical model that self-consistently describes the spectro-polarimetric properties of the thermal emission. We replaced {\tt kerrbb} with the model {\tt kynbbrr}. Similar to {\tt kerrbb}, {\tt kynbbrr} assumes a standard \cite{Novikov1973} disc model. However, it provides descriptions for both the spectral and polarimetric properties of the emitted radiation.
The disc local radiation is assumed to be polarised as in an electron-scattering dominated semi-infinite atmosphere \citep{Chandrasekhar1960,Sobolev1963}. This model is an extension of the relativistic package \textsc{kyn} \citep{Dovciak+04,Dovciak+08} developed to include the contribution of returning radiation \citep{Schnittman+09,Taverna+20}. The returning radiation contribution is regulated by the {\tt albedo} parameter, which describes the fraction of returning photons reflected from the disc surface while the remainder are absorbed. For a more detailed theoretical description of the model, we refer to \citet{Mikusincova+23} and references therein. 

\begin{table}
\begin{center}
\resizebox{\columnwidth}{!}{
\begin{tabular}[t]{ccc}
\hline
\hline

Component & Parameter (unit) & Values \\
\hline

{\tt kerrbb} & Mass accretion rate ($10^{16} \mathrm{g \ s^{-1}}$) & $3.54 \pm 0.03$ \\
& norm ($10^{-3}$) & $5.67 \pm 1.38$ \\

& $\chi^2$/dof & $436/442$\\
\hline

{\tt kynbbrr} & Mass accretion rate ($ 10^{-2} \ \mathrm{M_{Edd}}$) & $1.64^{+0.02}_{-0.05}$ \\
 & norm & $1.92 \pm 0.03$ \\

& $\chi^2$/dof & $488/445$\\
\hline
{\tt kynbbrr} & $\chi_0$ ($^\circ$) & $-67.5^{+22.9}_{-12.5}$ \\
 & albedo & Unconstrained \\
{\tt polconst} & PD & Unconstrained \\
 & PA & Unconstrained \\

& $\chi^2$/dof & $73.4/62$\\
\hline
  
\end{tabular}
}
\end{center}
\caption{Best-fitting parameters obtained in the \ixpe spectral fit using Model \ref{eq:SpectralFit} and in the spectro-polarimetric analysis of \ixpe data using the {\tt kynbbrr} model, both detailed in Sect. \ref{sec:Pol_Fit}.}
\label{Tab:Spectro-pol_FIT}
\end{table}

\begin{figure}
    \centering
    \includegraphics[width=0.85\columnwidth]{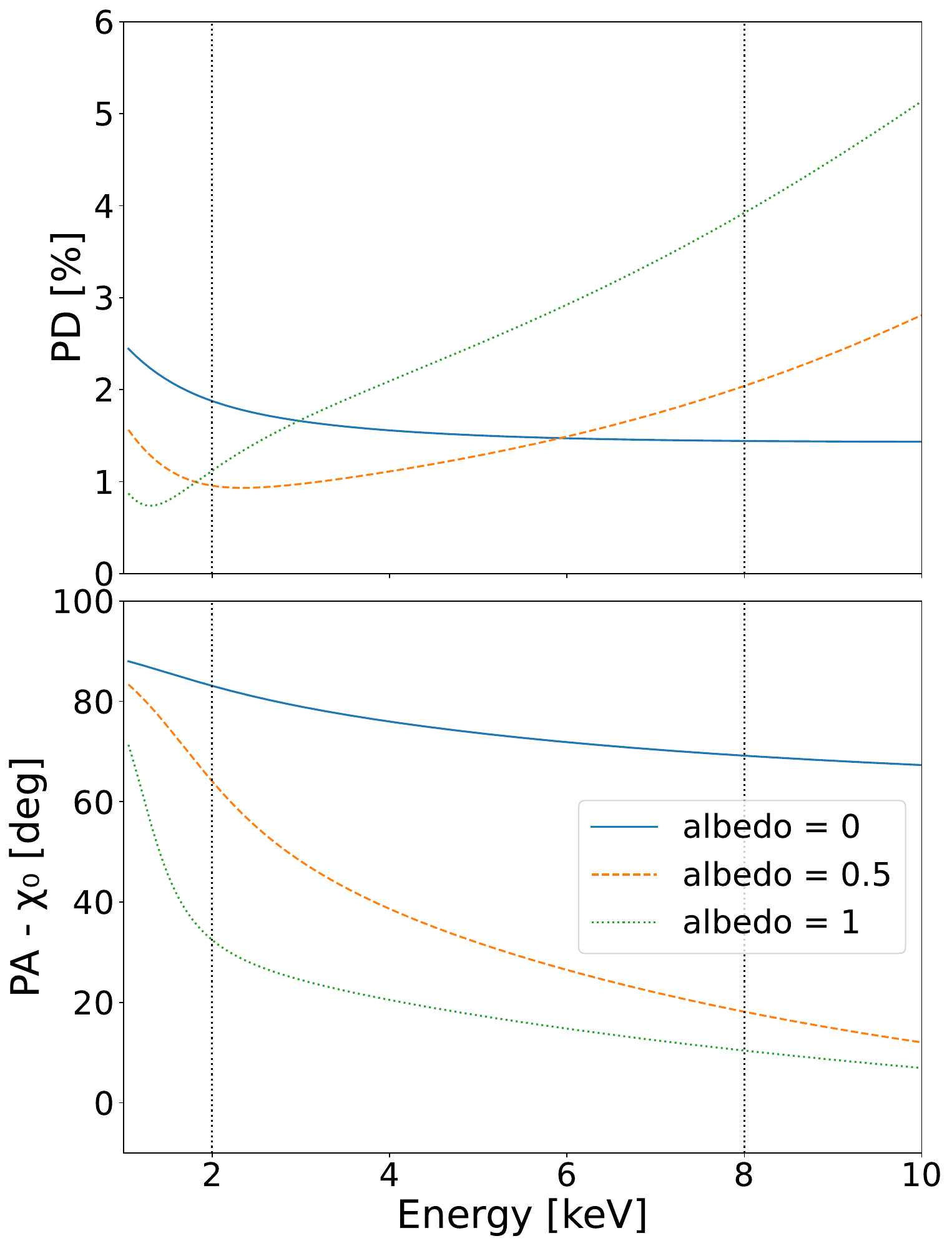}
    \caption{PD (\emph{top}) and PA (\emph{bottom}) predicted by the {\tt kynbbrr} model for the disc emission, assuming all system parameters are fixed at their respective best-fitting values (see Table \ref{Tab:SpectralFit}). The different colours indicate different contributions of the returning radiation component, regulated by the {\tt albedo} parameter. The vertical dashed lines highlight the 2--8 keV energy range. Here, the PA is defined with respect to the disc axis position angle $\chi_0$.}
    \label{fig:KYN_BestFit}
\end{figure}

\begin{figure}
    \centering
    \includegraphics[width=\columnwidth]{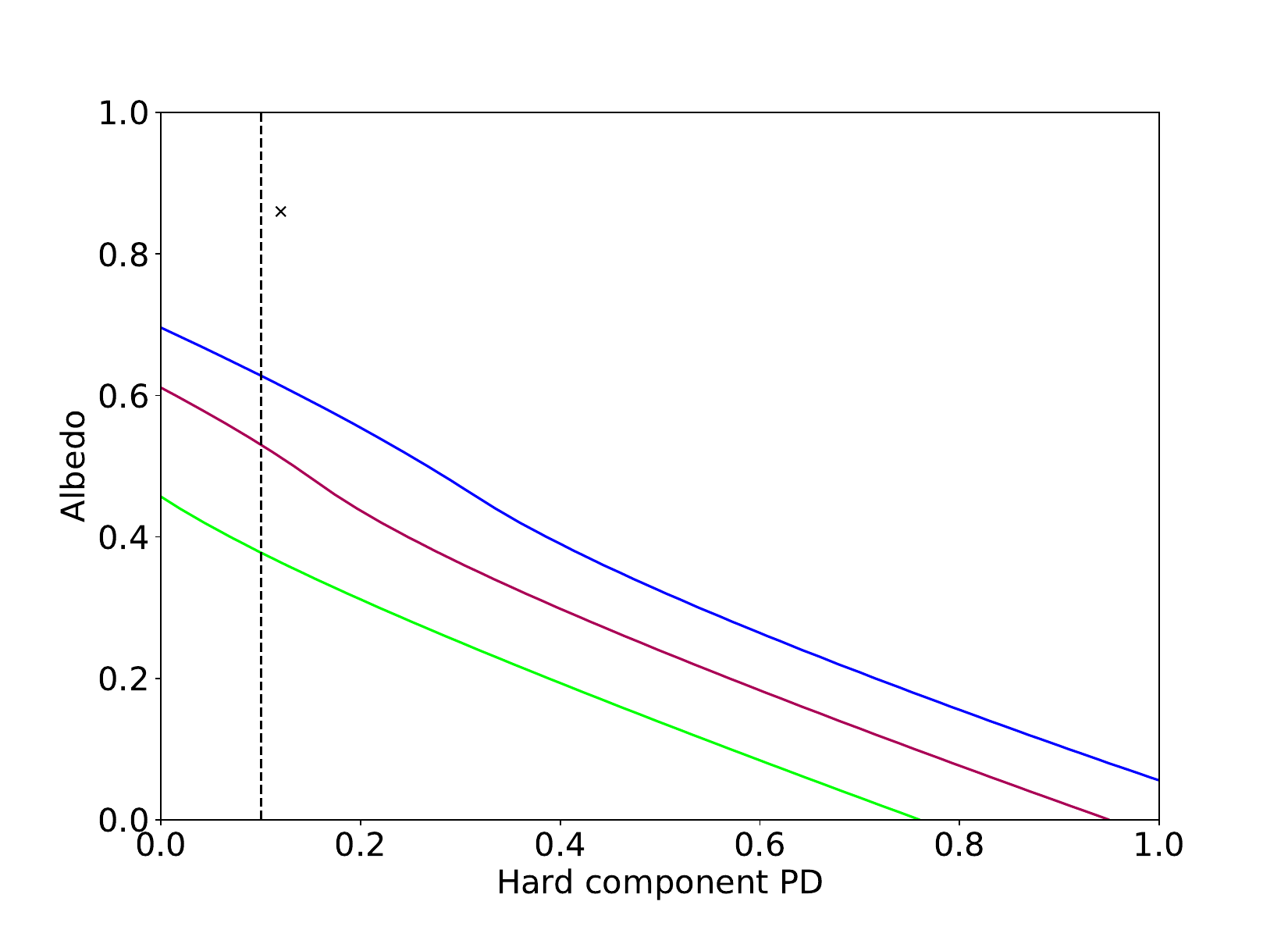}
    \caption{Contour plot of the corona emission PD and the {\tt kynbbrr} {\tt albedo} parameter, which regulates the returning radiation contribution. Blue, red, and green lines indicate 68\%, 90\%, and 99\% confidence levels for two parameters of interest, respectively, while the black cross indicates the best-fitting parameters. The dotted vertical line represents the assumed upper limit on the corona emission PD, as described in Sect. \ref{sec:Pol_Fit}.}
    \label{fig:CPlot_BestFit}
\end{figure}

To implement this model in the analysis we first used it to fit the \ixpe $I$ spectra, leaving only the accretion rate and the normalisation as free parameters. The best-fitting values for these parameters are shown in Table \ref{Tab:Spectro-pol_FIT}. The fit resulted in $\chi^2$/dof=488/445, assuming no contribution from the returning radiation ({\tt albedo}$=0$); the resulting spectral fit was insensitive to variation of the {\tt albedo} parameter. This absence of a spectral signature attributed to the returning radiation component deviates from recent findings in soft state BH outbursts. In these contexts, returning radiation has been proposed as a plausible source for observed relativistic reflection features, as seen in studies like \citep{Connors+20,Connors+21,Lazar+21}. These investigations, alongside theoretical predictions by \cite{Dauser+22}, characterised the returning radiation component using the {\tt relxillNS} model, an evolution of the {\tt relxill} suite of relativistic reflection models that assumes a single-temperature black-body as the incident radiation source. The discrepancy in results likely arises from the crucial difference in the way reflection is treated in the two models: while {\tt kynbbrr} describes the reflection process using \cite{Chandrasekhar1960} diffuse reflection formulae, presuming a completely ionised disc atmosphere \cite[see also][]{Taverna+20}, it lacks the capacity to replicate any reflection features in the spectra, as {\tt relxillNS} does. Since we have not found any apparent reflection features present in the spectra, we consider the use of this pure scattering approximation in our spectro-polarimetric analysis justified. This choice is reinforced by the \cite{Taverna+21} results, which indicate that, due to higher temperatures and lower plasma densities, the matter within the inner regions of rapidly rotating BHs accretion discs is anticipated to be almost entirely ionised.

As the parameters obtained were consistent with the values obtained with Model \ref{eq:SpectralFit}, we extracted from the code the theoretical prediction for the thermal emission PD and PA, presented in Fig.~\ref{fig:KYN_BestFit}. As the self-irradiation contribution becomes more significant, the PD shows an increase with energy due to the large PD expected for this component. Simultaneously, the PA exhibits a $90\degr$ rotation with energy, as the returning photons are expected to be polarised perpendicularly to the ones that directly reach the observer after leaving the disc atmosphere. If the albedo parameter is large enough, this rotation occurs below $2$ keV, leading to a relatively constant behaviour of the PA with energy in the \ixpe energy interval \citep{Taverna+20}.
Subsequently, we froze all the spectral parameters of the model and focused on the fit of \ixpe $Q$ and $U$ spectra. We employed a {\tt polconst} model to describe the hard component polarisation properties and left its parameters free to vary in the fit together with the {\tt kynbbrr} {\tt albedo} and orientation ($\chi_0$) parameter, which indicates the accretion disc axis position angle. We obtained a $\chi^2$/dof=73.4/62, for the best-fit values detailed in Table \ref{Tab:Spectro-pol_FIT}. However, the soft component {\tt albedo} and the hard component PD and PA remained unconstrained during the fitting procedure. This can be understood by looking at the contour plots presented in Fig.~\ref{fig:CPlot_BestFit}, which shows the 68\%, 90\%, and 99\% confidence level contours for the allowed values of the hard component PD and the {\tt albedo} parameter. The contours indicate a degeneracy between the two parameters, suggesting two different ways to explain the increasing trend of the PD with energy: either a very large PD of the hard component or a strong contribution from returning radiation. 

\begin{figure*}
    \centering
    \includegraphics[width=\textwidth]{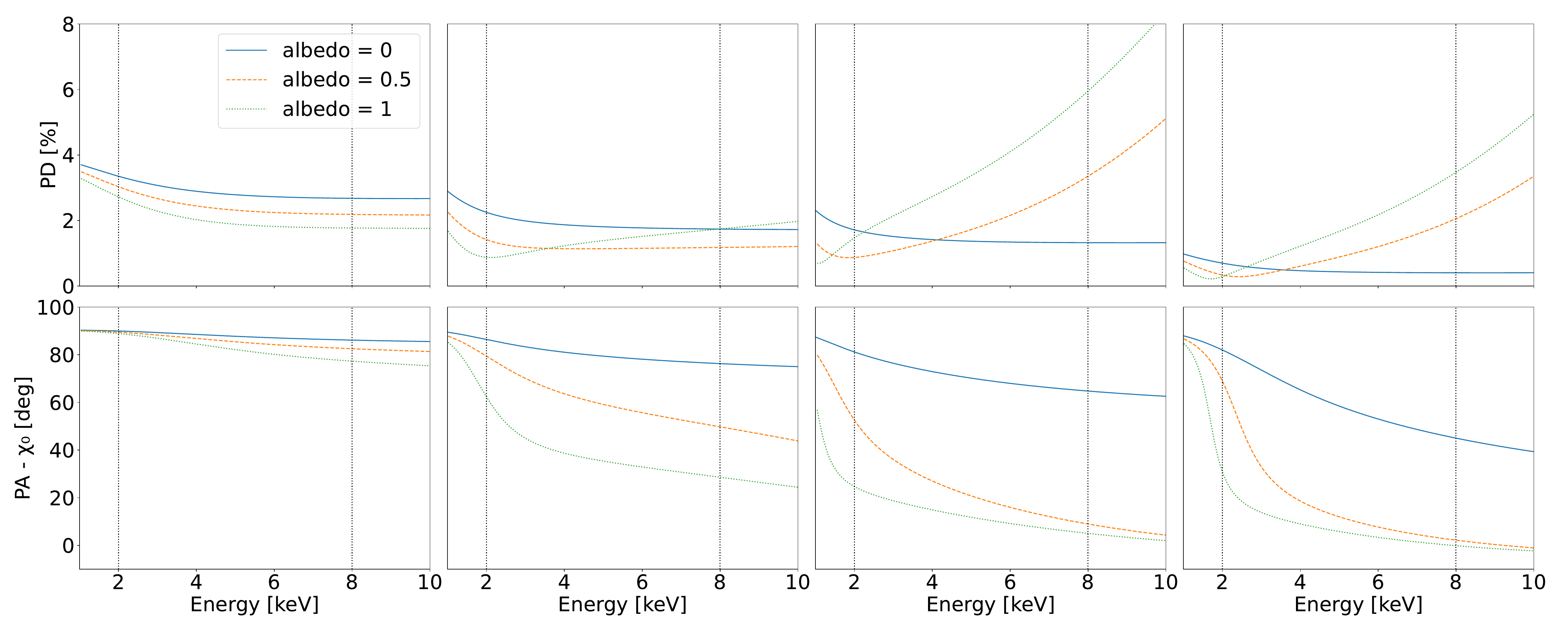}
    \caption{PD (\emph{top row}) and PA (\emph{bottom row}) predicted by the \texttt{kynbbrr} model for the disc emission assuming different values for the BH spin and for the system inclination (from left to right): $a = 0.5$ and $i=75\degr$, $a = 0.96$ and $i=75\degr$, $a = 0.998$ and $i=75\degr$, and $a = 0.998$ and $i=50\degr$. The different colours indicate different contributions of the returning radiation component, regulated by the {\tt albedo} parameter. The vertical dashed lines highlight the 2--8 keV energy range. 
    Here, the PA is defined with respect to the disc axis position angle $\chi_0$.}
    \label{fig:PolMod_KYN}
\end{figure*}

The polarisation properties of coronal emission are influenced by various factors, such as its geometry, optical depth, location, and velocity \citep[see e.g.][]{Zhang+22}. Assuming a flat corona geometry, theoretical studies suggest that the polarisation vector aligns with the disc axis \citep{Poutanen+96,Schnittman+10,KrawczynskiBeheshtipour+22}. This is indeed what has been found in the \ixpe observation of Cyg X-1 in the hard state \citep{Krawczynsky+22}. This alignment results in the polarisation vector being parallel to that of the returning radiation and perpendicular to the disc direct emission \citep{Schnittman+09,Taverna+20}. Additionally, the \ixpe observation of 4U~1630$-$47 in the steep power-law state measured a PD of the coronal emission of about 7\% \citep{RodriguezCavero+23}. Considering the similarities between these two sources, both being accreting BH systems likely observed at large inclination angles, we imposed an upper limit of 10\% on the PD of the hard component of 4U~1957+115. With this assumption, the only viable explanation for the polarimetric data is the inclusion of returning radiation. Hence, our polarimetric fit shows that the standard thin disc model can effectively describe the polarimetric data of this source, but it is necessary to consider the contribution from self-irradiation (assuming a 10\% PD for the corona emission, we find a lower limit of $0.73$ for the {\tt albedo} parameter).

As detailed in Sect.~\ref{sec:Pol_Data}, the initial part of our \ixpe observation did not yield a detectable polarisation signal, as outlined in Table~\ref{Tab:PD_PA}. Notably, our first {\it NuSTAR} observation, which displays the largest hard component contribution to the spectra, occurred near the end of this period. Considering the reduced hard component contribution during the rest of the observation, the observed low PD might be explained by depolarisation of radiation from the accretion disc by the corona emission. A similar situation was observed in LMC~X-1, where the low PD detected by \ixpe was attributed to the combination of two spectral components, disc and corona emission, polarised perpendicularly to each other \citep{Podgorny+23}. To investigate this scenario, we attempted to independently fit the polarimetric data in the first period. We made the assumption that the polarisation characteristics of the thermal emission remained constant throughout the observation and represented them using the best-fitting {\tt kynbbrr} model, derived from our analysis of the entire \ixpe observation. Employing a {\tt polconst} component to characterise the polarisation properties of the hard component, we estimate an upper limit on the PD of 17\%  during the initial phase of the \ixpe observation when assuming the corona emission to be polarised in the direction of the disc axis. This upper limit increased to 38\% in the perpendicular configuration. 

\begin{figure*}
    \centering
    \includegraphics[width=\textwidth]{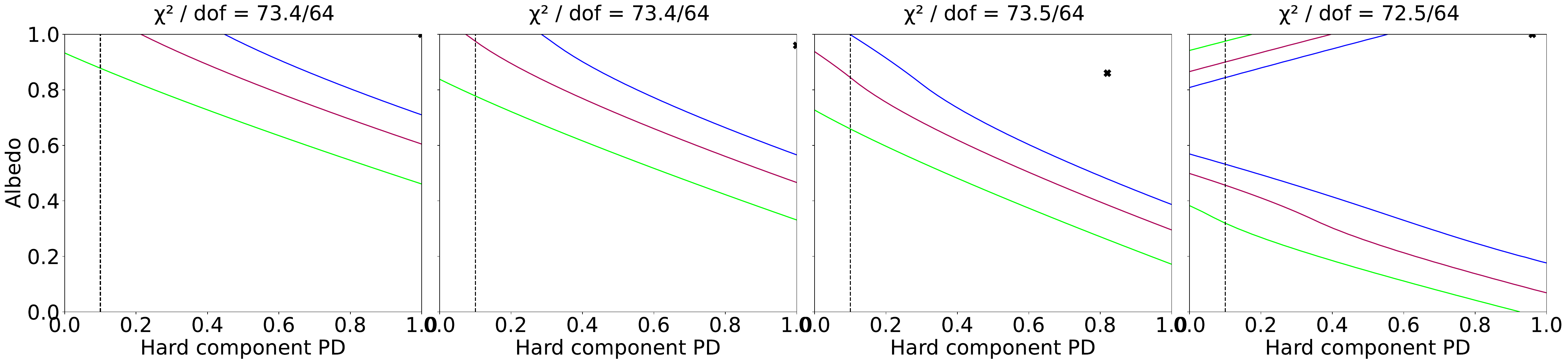}
    \caption{Contour plot of the corona emission PD and the \texttt{kynbbrr} {\tt albedo} parameter, assuming four different spin values (from left to right): $a = 0.95, 0.96, 0.97,$ and 0.998. The system inclination is assumed to be $75\degr$ in all cases. Blue, red and green lines indicate 68\%, 90\%, and 99\% confidence levels for two parameters of interest, respectively, while the black cross indicates the best-fitting parameters. The dotted vertical line represents the assumed upper limit on the corona emission PD, as described in Sect. \ref{sec:Pol_Fit}.}
    \label{fig:CPlot_Allspin}
\end{figure*}

\begin{figure*}
    \centering\includegraphics[width=\textwidth]{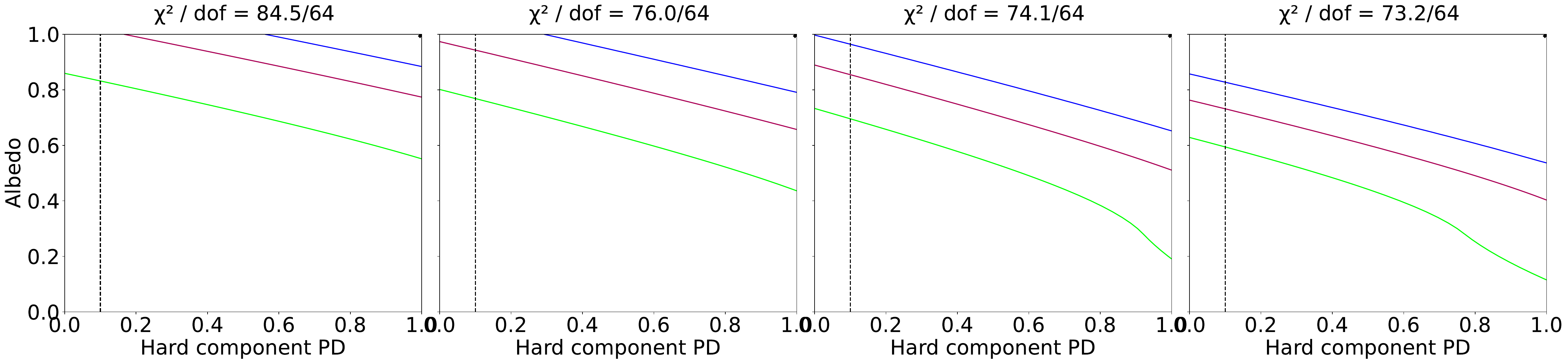}
    \caption{Same as Fig. \ref{fig:CPlot_Allspin}, but assuming a fixed spin value of 0.998 and considering different values for the accretion disc inclination (from left to right): $i=45\degr, 50\degr, 55\degr$, and 60\degr.}
    \label{fig:CPlot_Allinclinations}
\end{figure*}

\section{Spin and inclination constraints}\label{sec:SpinIncCost}\label{sec:KYNBBRR_spin}

Polarimetric data are known to be highly sensitive to the space-time of the emitting region. This is particularly true when considering accreting BHs in soft states because the dominant component of the X-ray spectra originates from photons emitted in the inner regions of the accretion disc, where the strong gravity around the central object significantly alters the observed polarisation properties \citep{Connors+77,Stark+77,Connors+80}. General relativity effects cause the photon polarisation vectors to undergo rotation as they propagate along geodesics, resulting in a net depolarisation of the emission at infinity and an overall rotation of the PA. Additionally, the already mentioned contribution of returning radiation can profoundly impact the polarimetric signature of the observed radiation \citep{Schnittman+09}.
These effects are expected to be stronger for the high-energy photons emitted closer to the central BH, introducing an energy dependence on the observed polarisation properties of the thermal emission that provides a mean to estimate the extent of the accretion disc near the BH. In soft-state sources, it is widely believed that the accretion disc extends down to the innermost stable circular orbit (ISCO; \citealt{TanakaLewin, Steiner+2010}), making polarimetric data a valuable tool for estimating the BH spin in such sources via the well-known relation between the ISCO radius and spin \citep[e.g.][]{Fabian+19}. Furthermore, the inclination of the accretion disc also significantly influences the observed polarimetric properties, making modelling of polarimetric data crucial in extracting information about this parameter \citep{Dovciak+08,Taverna+20}.

The polarimetric fit described in Sect. \ref{sec:Pol_Fit} indicates that the standard Novikov-Thorne thin disc model \citep{Novikov1973,Kato2008}, assuming the large spin and inclination values found in the spectral analysis described in Sect. \ref{Sec:SpectralAnalysis}, successfully explains the polarimetric data. Here, our goal is to test whether the standard thin disc model can effectively describe the polarimetric data for different values for the BH spin and the disc inclination.

To model the polarimetric data with \texttt{kynbbrr}, we repeated the procedure described in Sect. \ref{sec:Pol_Fit}, but this time we explored different values for the BH spin and the system inclination. The spectral fit favoured configurations with large spin and inclination values, so when we reduced these parameters' values, we had to relax some of the initial assumptions on the model to achieve an acceptable fit. Specifically, we allowed the source distance and disc hardening factor to vary freely, as they tend to increase significantly when considering lower spin and inclination values, albeit above the limits suggested by \textit{Gaia} parallax measurements \citep{Maccarone+20,Barillier+23} and disc atmosphere modelling \citep{Shimura+95}. 

Figure \ref{fig:PolMod_KYN} displays the soft component PD and PA predicted by the \texttt{kynbbrr} fit for different spin and inclination values. As the BH spin decreases, the ISCO location moves further away from the central BH. Consequently, the fraction of photons forced to return to the disc surface diminishes.  When the BH spin becomes sufficiently low ($a \lesssim 0.96$), the contribution of returning radiation is no longer sufficient to explain the increase in PD with energy, and its primary effect is to depolarise the direct emission. On the other hand, when the disc inclination angle decreases, the observed PD decreases across the entire \ixpe band, while the PA exhibits larger rotations with energy.
We conducted a polarimetric fit on the \ixpe $Q$ and $U$ spectra while assuming different values for the BH spin and the disc inclination. Our analysis identified that the configuration with spin $a = 0.998$ and disc inclination $i=75\degr$ provided the most accurate description of the data, resulting in a $\chi^2$/dof of $72.5/62$. On the other hand, the best polarimetric fit assuming spin $a=0.5$ resulted in the notably larger $\chi^2$/dof $=79.2/62$; furthermore, this configuration demands an exceptionally high PD for the corona component, with a determined lower limit of $51\%$. When assuming the fiducial value of $10\%$, the $\chi^2$/dof increases to $84.9/63$. This considerably worse fit results from the absence of a significant increase in energy of the PD within the \ixpe energy band, as illustrated in the leftmost column of Figure \ref{fig:PolMod_KYN}. Figures \ref{fig:CPlot_Allspin} and \ref{fig:CPlot_Allinclinations} present the chi-squared values and contour plots between the PD of the corona emission and the \texttt{kynbbrr} albedo parameter for four different combinations of these parameters. As the BH spin or the disc inclination angle decreases, a larger corona PD is required to account for the PD increase with energy. Notably, for cases where $a < 0.96$ or when assuming a disc inclination lower than $50\degr$, the corona PD exceeds the upper limit of 10\% that we discussed in Sect. \ref{sec:Pol_Fit}. Consequently, assuming a standard thin disc model, these configurations are disfavoured.

\section{Returning radiation and corona optical depth predictions with \texttt{kerrC}}
\label{sec:kerrC}

In order to ascertain the effect that returning radiation, along with the presence or absence of a corona, has on the observed polarisation properties of the source, we employed the general relativistic ray-tracing code \texttt{kerrC} \citep{KrawczynskiBeheshtipour+22, Krawczynsky+22}. The code operates under the prescription of a standard geometrically thin, optically thick accretion disc spanning from the ISCO to 100 gravitational radii and draws from a library of 68,040 BH, accretion flow, and corona configurations. Using the NICER and {\it NuSTAR} joint fit described in Sect. \ref{Sec:SpectralAnalysis} we fixed the absorption, BH mass, distance, spin, mass accretion rate, and inclination to the values in Table \ref{Tab:SpectralFit} corresponding to Period 2. We used {\tt kerrC} to simulate the polarisation of the source for the cases in which returning radiation is either disregarded or considered towards the total emission. The coronal optical depth $\tau_{\rm C}$ is measured vertically from the accretion disc to the upper edge of the corona. A value of $\tau_{\rm C} = 0$ corresponds to a system where no coronal plasma is present and the emission consists only of direct emission from the disc and, optionally, of returning emission owing to space-time curvature without scattering in the corona. As the coronal optical depth increases, the total emission includes reflected emission from the disc onto the corona, which increases the PD \citep{Schnittman+09}.

In our predictions, we used a wedge-shaped corona with a temperature of 100 keV. The corona opening angle was $10\degr$, the photospheric electron density was set to $\log_{10} n_{\rm e} = 17.5,$ and the metallicity to $A_{\rm Fe} = 1.0$ as per \citet{Krawczynsky+22}.
We employed three different corona optical depths, $\tau_{\rm C} = $ 0, 0.01, and 0.05, at the proposed inclination of $75\degr$.
In the case where returning radiation is neglected, the predicted PD increases with the presence of a corona: from 1.88\% without a corona to 2.02\% and 2.01\% with corona optical depths of 0.01 and 0.05, respectively, in the 2--8 keV band. While these PDs are well within the confidence region of the observation, they fail to replicate the rise of PD with respect to energy shown in Figure \ref{fig:PD-PA-vsE}. For optical depths 0.0 and 0.01, the PD decreases by 0.69\% and 0.32\%, respectively, across the \ixpe band. When the corona optical depth is 0.05, the PD only experiences a slight increase of 0.04\%.

In the case where returning radiation is taken into consideration, the predicted PDs are 3.50\%, 3.28\%, and 3.43\% for corona optical depths of 0.0, 0.01, and 0.05. The predicted 2-8 keV energy-averaged PD is higher than the observed PD, which may be attributed to the 6-8 keV PD being below the MDP. When returning radiation is considered, the PD increases from 1.54\%, 1.60\%, and 1.78\%  between 2 and 3 keV to 2.89\%, 3.04\%, and 3.59\% in the 4.3--6.0~keV energy band in accordance to the observation.
Our simulations with {\tt kerrC} suggest that returning radiation is necessary to reproduce the \ixpe polarisation results.
Additionally, we tested a spin of $a = 0.95$ for the configurations with the highest energy average PD, namely for the case of returning radiation and a corona optical depths 0.0 and 0.05.
For $\tau_C = 0.0$, the PD is 1.18\% between 2 and 3 keV and increases to 1.41\% between 4.3 and 6.0 keV.
For $\tau_C = 0.05$, the PD increases from 1.15\% to 1.42\% in the same energy bands.
A {\tt kerrC} model of low spin fails to reproduce the data.
A comprehensive examination of the source's spectro-polarimetric characteristics and partial contributions of returning radiation using the \texttt{kerrC} code exceeds the scope of this study and will be addressed in a future work.


\section{Conclusions}\label{sec:Conclusions}

We performed an X-ray spectro-polarimetric observational campaign on the accreting low-mass X-ray binary BH system 4U~1957+115, with coverage by the \ixpe, NICER, {\it NuSTAR}, and SRG missions. Our spectral analysis indicates that the source was in a soft state, characterised by a predominant thermal disc emission, with only a minor contribution from a Comptonised component and no clear reflection features. We observed a diminishing trend in the contribution of the hard X-ray tail during the observation period, with the initial {\it NuSTAR} data exhibiting the  strongest power-law tail. Notably, this trend was not discernable from the \ixpe and NICER data, highlighting that the Comptonisation component becomes relevant only above $\sim$10 keV and contributes only marginally in the \ixpe energy range (2.3\%, 1\%, and 0.7\% during the three {\it NuSTAR} observations). Our spectral fitting, which used a relativistic accretion disc emission model, strongly favours configurations characterised by large inclination angles and high spin values. By fixing the inclination, mass, and distance parameters to fiducial values, we estimate the BH spin to be $0.992 \pm 0.003$, which aligns with the literature values, considering the uncertainties \citep[but see][]{Sharma+21}. The polarimetric observation of 4U~1957+115 revealed a time-averaged 2--8~keV PD of $1.9\% \pm 0.6\%$ with a PA of $-41\fdg8 \pm 7\fdg9$. The PA remains constant across different energy bins, while the PD shows a slight increase with energy, rising from $\approx 1.6$\% between 2 and 3 keV to $\approx 3.1$\% in the 4.3--6.0~keV energy band. The observed polarimetric data are consistent with theoretical predictions for thermal emission originating from an optically thick and geometrically thin disc with a Novikov-Thorne profile, assuming \citet{Chandrasekhar1960} and \citet{Sobolev1963} prescriptions for polarisation due to electron scattering in semi-infinite atmospheres. This agreement is achieved by accounting for the substantial contribution of self-irradiation.
It is important to note that the inclusion of absorption effects could alternatively give rise to an increase in the PD with energy, mimicking the contribution of returning radiation \citep{Ratheesh+23}. The impact of absorption, however, is estimated to be negligible in the disc atmosphere of rapidly rotating BHs, because matter in the inner regions of the accretion disc is expected to be almost completely ionised due to the higher temperatures and lower densities of the plasma \citep{Taverna+21}. Additionally, due to the lack of spectral features, we can rule out any contribution of highly ionised gas along the line of sight
Therefore, we regard the pure scattering atmosphere as a reasonable approximation to model our data. 
Our spectro-polarimetric analysis indicates that configurations with low BH spin values or low inclination angles are disfavoured within the standard Novikov-Thorne thin disc model, in agreement with the spectral analysis. In fact, such configurations struggle to explain the observed increase in PD with energy without requiring unphysically high PD values for the power-law component. 

\section*{Acknowledgements}

The Imaging X-ray Polarimetry Explorer (IXPE) is a joint US and Italian mission.  The US contribution is supported by the National Aeronautics and Space Administration (NASA) and led and managed by its Marshall Space Flight Center (MSFC), with industry partner Ball Aerospace (contract NNM15AA18C).  The Italian contribution is supported by the Italian Space Agency (Agenzia Spaziale Italiana, ASI) through contract ASI-OHBI-2022-13-I.0, agreements ASI-INAF-2022-19-HH.0 and ASI-INFN-2017.13-H0, and its Space Science Data Center (SSDC) with agreements ASI-INAF-2022-14-HH.0 and ASI-INFN 2021-43-HH.0, and by the Istituto Nazionale di Astrofisica (INAF) and the Istituto Nazionale di Fisica Nucleare (INFN) in Italy.  This research used data products provided by the IXPE Team (MSFC, SSDC, INAF, and INFN) and distributed with additional software tools by the High-Energy Astrophysics Science Archive Research Center (HEASARC), at NASA Goddard Space Flight Center (GSFC).

M. Brigitte acknowledges the support from GAUK project No. 102323.
N.R.C. and H.K. acknowledge support by NASA grants 80NSSC22K1291, 80NSSC23K1041, and 80NSSC20K0329.
A.V. thanks the Academy of Finland grant 355672 for support.
M.P. and P.-O.P. acknowledge financial support from the High Energy National Programme (PNHE) of the Scientific Research National center (CNRS) and from the french spatial agency (Centre National d'Etudes Spatiales, CNES).
M.D., J.Pod., J.S. and V.K. thank for the support from the GACR project 21-06825X and the institutional support from RVO:67985815.
I.L. was supported by the NASA Postdoctoral Program at the Marshall Space Flight Center, administered by Oak Ridge Associated Universities under contract with NASA.
This research was also supported by the INAF grant 1.05.23.05.06: “Spin and Geometry in accreting X-ray binaries: The first multi frequency spectro-polarimetric campaign".



\bibliographystyle{aa} 
\bibliography{bh} 
\end{document}